# Evolution and mass extinctions as lognormal stochastic processes

Claudio Maccone

*International Academy of Astronautics (IAA), IAA SETI Permanent Committee, Istituto Nazionale di Astrofisica (INAF), Via Martorelli, 43 – Torino (Turin) 10155, Italy*
*e-mail: clmaccon@libero.it and claudio.maccone@iaamail.org*

**Abstract**: In a series of recent papers and in a book, this author put forward a mathematical model capable of embracing the search for extra-terrestrial intelligence (SETI), Darwinian Evolution and Human History into a single, unified statistical picture, concisely called *Evo-SETI*. The relevant mathematical tools are:

(1) Geometric Brownian motion (GBM), the stochastic process representing evolution as the stochastic increase of the number of species living on Earth over the last 3.5 billion years. This GBM is well known in the mathematics of finances (Black–Sholes models). Its main features are that its probability density function (pdf) is a lognormal pdf, and its mean value is either an increasing or, more rarely, decreasing exponential function of the time.

(2) The probability distributions known as *b*-lognormals, i.e. lognormals starting at a certain positive instant $b > 0$ rather than at the origin. These *b*-lognormals were then forced by us to have their peak value located on the exponential mean-value curve of the GBM (Peak-Locus theorem). In the framework of Darwinian Evolution, the resulting mathematical construction was shown to be what evolutionary biologists call *Cladistics*.

(3) The (Shannon) *entropy* of such *b*-lognormals is then seen to represent the 'degree of progress' reached by each living organism or by each big set of living organisms, like historic human civilizations. Having understood this fact, human history may then be cast into the language of *b*-lognormals that are more and more organized in time (i.e. having smaller and smaller entropy, or smaller and smaller 'chaos'), and have their peaks on the increasing GBM exponential. This exponential is thus the 'trend of progress' in human history.

(4) All these results also match with SETI in that the statistical Drake equation (generalization of the ordinary Drake equation to encompass statistics) leads just to the lognormal distribution as the probability distribution for the number of extra-terrestrial civilizations existing in the Galaxy (as a consequence of the central limit theorem of statistics).

(5) But the most striking new result is that the well-known 'Molecular Clock of Evolution', namely the 'constant rate of Evolution at the molecular level' as shown by Kimura's Neutral Theory of Molecular Evolution, *identifies* with growth rate of the entropy of our Evo-SETI model, because they both grew *linearly* in time since the origin of life.

(6) Furthermore, we apply our Evo-SETI model to lognormal stochastic processes *other than GBMs.* For instance, we provide two models for the mass extinctions that occurred in the past: (a) one based on GBMs and (b) the other based on a *parabolic* mean value capable of covering both the extinction and the subsequent recovery of life forms.

(7) Finally, we show that the Markov & Korotayev (2007, 2008) model for Darwinian Evolution identifies with an Evo-SETI model for which the mean value of the underlying lognormal stochastic process is a *cubic* function of the time.

In conclusion: we have provided a new mathematical model capable of embracing molecular evolution, SETI and entropy into a simple set of statistical equations based upon *b*-lognormals and lognormal stochastic processes with arbitrary mean, of which the GBMs are the *particular case of exponential growth.*

## Introduction: mathematics and science

Sir Isaac Newton published his law of universal gravitation in 1687. In the following 250 years ($\sim$ 1700–1950) many eminent mathematicians developed celestial mechanics (i.e. the theory of orbits) manually. The result of all those 'difficult calculations' was seen after the advent of computers in 1950: as of 2014 we have a host of spacecrafts of all types flying around and beyond the solar system: in other words, the space age is the outcome of both the law of universal gravitation and of lots of mathematics.

Similarly, between 1861 and 1862 James Clerk Maxwell first published 'the Maxwell equations'. These summarized all previous experimental work in electricity and magnetism and paved the mathematical way to all subsequent discoveries, from radio waves to cell phones. Again a bunch of equations was the turning point in the history of humankind, although very few people know about 'all those mathematical details' even in today's Internet age.

Having so said, this author believes that the time is ripe for a brand-new mathematical synthesis embracing the whole of Darwinian Evolution, human history and the search for extra terrestrial intelligence (SETI) into a bunch of simple equations. But these equations have to be *statistical*, rather than deterministic as it was in the case of Newton's and Maxwell's equations. In fact, the number of possible examples covered by these equations is huge and thus it may be handled only by virtue of statistics. Just imagine the number of Earth-type exoplanets existing in the Milky Way in 2014 is estimated to be around 40 billion. Then, if we are going to *predict* what stage in the evolution of life a certain newly discovered exoplanet may have reached, our predictions may only be *statistical*.

The Evo-SETI theory, outlined in this paper, is intended to be the correct mathematical way to let humans 'classify' any newly discovered exoplanet or even an alien civilization (as required by SETI) according to its own 'degree of evolution', given by the entropy of the associated *b*-lognormal probability density function (pdf), as we shall see in the section 'Entropy as the evolution measure' of this paper. Also, the index of evolution (Evo-Index) defined in the section 'Entropy as the evolution measure' measures the positive evolution starting from zero at the time of the origin of life on that exoplanet, and so we presume that we have found a way to *quantify progress in the evolution of life, from RNA to humans and on to extraterrestrials.*

## A summary of the 'Evo-SETI' model of evolution and SETI

This paper describes recent developments in a new statistical theory casting evolution and SETI into mathematical equations, rather than just using words only: this we call the Evo-SETI model of evolution and SETI. Our final goal is to prove that the Evo-SETI model and the well-known molecular clock of evolution are in agreement with each other. In fact, the (Shannon) entropy of the *b*-lognormals in the Evo-SETI model decreases *linearly* with time, just as the molecular clock increases *linearly* with time. Apart from constants with respect to the time, *b*-lognormals entropy and molecular clock are the same.

However, the calculations required to prove them are lengthy. To overcome this obstacle, the Appendix (available in the supplementary material) gives a summary of all the analytical calculations that this author performed by the Maxima symbolic manipulator especially to prove the Peak-Locus theorem described in section 'Peak-Locus theorem'. It is interesting to point out that the Macsyma symbolic manipulator or 'computer algebra code' (of which Maxima is a large subset) was created by NASA at the Artificial Intelligence Laboratory of MIT in the 1960s to check the equations of celestial mechanics that had been worked out manually by a host of mathematicians in the previous 250 years (1700–1950). Actually, those equations might have contained errors that could have jeopardized the Moon landings of the Apollo Program, and so NASA needed to check them by computers, and Macsyma (nowadays Maxima) did a wonderful job. Today, anyone can download Maxima *for free* from the website http://maxima.sourceforge.net/. The Appendix of this paper (available in the supplementary material) is written in Maxima language and the conventions applied for denoting the input instructions by (%i[equation number]) and the output results by (%o[equation number]), as we shall see shortly.

Going now back to the general lognormal stochastic process $L(t)$ (standing for 'life at $t$', and also for 'lognormal stochastic process at time $t$'), let us first point out that:

(1) $L(t)$ starts at a certain time $t = ts$ with certainty, i.e. with probability one. For instance, if we wish to represent the evolution of life on Earth as a stochastic process in the number of species living on Earth at a certain time $t$, then the starting time $t = ts$ will be the time of the origin of life on Earth. Although we do not know exactly when that occurred, we approximately set it at $ts = -3.5$ billion years, with the convention that past times are negative times, the present time is $t = 0$ and future times will be positive times.

(2) We now make the *basic mathematical assumption* that the stochastic process $L(t)$ is a *lognormal process starting at* $t = ts$, namely its pdf is given by the lognormal

$$L(t)_\mathrm{pdf}\,(n; M_L(t), \sigma_L, t) = \frac{e^{-\frac{[\ln(n) - M_L(t)]^2}{2\sigma_L^2 (t-ts)}}}{\sqrt{2\pi}\,\sigma_L \sqrt{t-ts}\,n} \tag{1}$$

with $\begin{cases} n > 0 \\ t \geqslant ts \end{cases}$ and $\begin{cases} \sigma_L \geqslant 0, \\ M_L(t) = \text{arbitrary function of } t. \end{cases}$

Profound 'philosophical justifications' exist behind the assumption summarized by equation (1): for instance, the fact that 'lognormal distributions are necessarily brought into the picture by the central limit theorem of statistics' (Maccone 2008, 2010a). However, we will not be dragged into this mountain of philosophical debates since this author is a mathematical physicist wanting to make progress in understanding nature, rather than wasting time in endless debates. Therefore, we now go on to our third basic assumption.

(3) The mean value of the process $L(t)$ is an *arbitrary* (and continuous) function of the time denoted by $m_L(t)$ in the sequel. In equations, that is, one has, by definition

$$m_L(t) = \langle L(t) \rangle. \tag{2}$$

In other words, we analytically compute the following integral, yielding the mean value of the pdf (1), getting (for proof, see (%o5) and (%o6) in the Appendix available in the supplementary material)

$$m_L(t) \equiv \int_0^\infty n \frac{e^{-\frac{[\ln(n)-M_L(t)]^2}{2\sigma_L^2(t-ts)}}}{\sqrt{2\pi}\,\sigma_L\sqrt{t-ts}\,n}\,dn = e^{M_L(t)}e^{\frac{\sigma_L^2}{2}(t-ts)}. \tag{3}$$

There are two functions of the time in (3): $m_L(t)$ and $M_L(t)$. Since we assumed $m_L(t)$ to be an arbitrary continuous function of the time, it follows from (3) that $M_L(t)$ must also be so, and they may be freely interchanged, since (3) may be exactly solved with respect to either of them. Also, please do not worry about 'the variance $\sigma_L$ of $L(t)$' in (3): we shall shortly see at point 7) that $\sigma_L$ is determined by both the arbitrary mean value function $m_L(t)$ and the standard deviation of the process $L(t)$ at the end time $te$ by $\delta Ne = \Delta(te)$. At this point, knowing the pdf (1) and the mean value (3), it is just a mathematical exercise to derive all the statistical properties of the stochastic process $L(t)$. Doing so, however, would require several pages of lengthy calculations (even by Maxima) that cannot be compressed in this paper. Thus, this author recommends readers to have a look at his recently published papers and his book (Maccone 2010a, 2011b, 2012, 2013), where the calculations yielding the statistical properties of $L(t)$ are described more in detail. Here, we have just summarized them in Table 1.

(4) At this point, the vertical axis of the $(t, L(t))$ plot still is undermined up to an arbitrary multiplicative constant. At the initial instant $ts$, we may thus denote by $Ns$ ('number at start') the numerically certain (i.e. with probability 1) starting value of the stochastic process $L(t)$. A few easy steps from (3) then show that, introducing $Ns$, the mean value (3) must be replaced by the 're-normalized' one

$$m_L(t) = Ns\; e^{M_L(t)-M_L(ts)}e^{\frac{\sigma_L^2}{2}(t-ts)} \tag{4}$$

(for the proof of this result, see (%i8) through (%o13) in the Appendix available in the supplementary material). In fact, if you set $t = ts$ into (4), both exponentials become 1, and one just gets the initial condition

$$m_L(ts) = Ns. \tag{5}$$

One may say that the two *assumed* numeric values $(ts, Ns)$ are the *initial boundary conditions* of the stochastic process $L(t)$.

(5) Similarly, one must specify the two numbers $(te, Ne)$ ('end time' and 'number at end time') representing the *final boundary conditions*, with

$$Ne = Ns\, e^{M_L(te)-M_L(ts)}e^{\frac{\sigma_L^2}{2}(te-ts)} \tag{6}$$

(see (%o23) in the Appendix available in the supplementary material).

(6) Replacing the mean value (3) by virtue of the re-normalized mean value (4) would lead to a new Table similar to Table 1 that, however, would contain the new term Ns. That new Table we will skip for the sake of brevity.

(7) The initial (5) and final (6) conditions only affect the mean value curve (4). That is not enough, however, since every stochastic process is determined not only by its mean value, but also by its two *upper and lower standard deviation curves* inferred from the higher moments of the lognormal pdf (1) and from (3), i.e. (see (%i14) through (%o21) in the Appendix available in the supplementary material)

$$\Delta_L(t) = e^{M_L(t)}e^{\frac{\sigma_L^2}{2}(t-ts)}\sqrt{e^{\sigma_L^2(t-ts)}-1} \tag{7}$$

Calling $\delta Ne = \Delta_L(te)$ the *final standard deviation* (namely the standard deviation at the end time $te$) this becomes one more input that must be assigned in addition to the four boundary conditions $(ts, Ns)$ and $(te, Ne)$ plus the arbitrary function $M_L(t)$. Then from (6) one may derive the promised $\sigma_L$ (see (%i22) through (%o27) in the Appendix available in the supplementary material)

$$\sigma_L = \frac{\sqrt{\ln\left[e^{2M_L(ts)} + (\delta Ne)^2\left(\frac{Ns}{Ne}\right)^2\right] - 2M_L(ts)}}{\sqrt{te-ts}} \tag{8}$$

to be inserted into the lognormal pdf (1), which is thus completely determined by the $M_L(t)$ arbitrary function plus the five numbers ($ts$, $Ns$, $te$, $Ne$ and $\delta Ne$). This completes the description of the $L(t)$ process.

## Important special cases of $m_L(t)$

(1) The particular case of (3) where the mean value is given by the generic exponential

$$m_{\rm GBM}(t) = N_0 e^{\mu_{\rm GBM}\,t} \text{ or, more easily, } = A\,e^{Bt} \tag{9}$$

is called geometric Brownian motion (GBM), and is widely used in financial mathematics, where it is the 'underlying process' of the stock values (Black–Scholes models (1973), or Black–Scholes–Merton models, with the Nobel prize in Economics awarded in 1997 to Sholes and Merton only since Black had unfortunately passed away in 1995). This author used the GBM in his previous mathematical models of evolution and SETI (Maccone 2010a, 2010b, 2011a, 2011b, 2012, 2013), since it was assumed that the growth of the number of ET civilizations in the Galaxy, or, alternatively, the number of living species on Earth over the last 3.5 billion years, *grew exponentially* (Malthusian growth). Notice also that, upon equating the two right-hand sides of (3) and (9), we find that

$$e^{M_{\rm GBM}(t)}\,e^{\frac{\sigma_{\rm GBM}^2}{2}(t-ts)} = N_0\,e^{\mu_{\rm GBM}(t-ts)}. \tag{10}$$

Solving this equation for $M_{\rm GBM}(t)$ yields

$$M_{\rm GBM}(t) = \ln N_0 + \left(\mu_{\rm GBM} - \frac{\sigma_{\rm GBM}^2}{2}\right)(t-ts). \tag{11}$$

Table 1. *Summary of the properties of the lognormal distribution that applies to the stochastic process $L(t)$ = lognormally changing number of ET communicating civilizations in the Galaxy, as well as the number of living species on Earth over the last 3.5 billion years. Clearly, these two different $L(t)$ lognormal stochastic processes may have two different time functions for $M_L(t)$ and two different numerical values for $\sigma_L$, but the equations are the same for both processes, i.e. for the number of ET civilizations in the Galaxy and for the number of living species in the past of Earth. This is the general lognormal growth, not necessarily Malthusian*

| | |
|---|---|
| Stochastic process | $L(t)=\begin{cases}(1)\ \text{Number of ET civilizations (in SETI)}\\(2)\ \text{Number of living species (in evolution)}\end{cases}$ |
| Probability distribution | Lognormal distribution of all lognormal stochastic processes, i.e. the lognormal stochastic processes with arbitrary mean $M_L(t)$ |
| Probability density function | $L(t)_\text{pdf}(n;M_L(t),\sigma_L,t)=\dfrac{e^{-\frac{[\ln(n)-M_L(t)]^2}{2\sigma_L^2(t-ts)}}}{\sqrt{2\pi}\,\sigma_L\sqrt{t-ts}\,n}$ with $\begin{cases}n>0\\t\geqslant ts\\\sigma_L\geqslant 0\\M_L(t)=\text{arbitrary function of } t\end{cases}$ |
| Mean value | $\langle L(t)\rangle \equiv m_{L(t)}=e^{M_L(t)}e^{\frac{\sigma_L^2}{2}(t-ts)}$ |
| Variance | $\sigma_{L(t)}^2=e^{2M_L(t)}e^{\sigma_L^2(t-ts)}\left(e^{\sigma_L^2(t-ts)}-1\right)$ |
| Standard deviation | $\sigma_{L(t)}=e^{M_L(t)}e^{\frac{\sigma_L^2}{2}(t-ts)}\sqrt{e^{\sigma_L^2(t-ts)}-1}$ |
| Upper standard deviation curve | $m_{L(t)}+\sigma_{L(t)}=e^{M_L(t)}e^{\frac{\sigma_L^2}{2}(t-ts)}\left(1+\sqrt{e^{\sigma_L^2(t-ts)}-1}\right)$ |
| Lower standard deviation curve | $m_{L(t)}-\sigma_{L(t)}=e^{M_L(t)}e^{\frac{\sigma_L^2}{2}(t-ts)}\left(1-\sqrt{e^{\sigma_L^2(t-ts)}-1}\right)$ |
| All the moments, i.e. $k$th moment | $\langle L^k(t)\rangle=e^{kM_L(t)}e^{(k^2-k)\frac{\sigma_L^2}{2}(t-ts)}$ |
| Mode (=abscissa of the lognormal peak) | $n_\text{mode}\equiv n_\text{peak}=e^{M_L(t)}e^{-\sigma_L^2(t-ts)}$ |
| Value of the mode peak | $f_{L(t)}(n_\text{mode})=\dfrac{e^{-M_L(t)}e^{\frac{\sigma_L^2}{2}(t-ts)}}{\sqrt{2\pi}\,\sigma_L\sqrt{t-ts}}$ |
| Median (=fifty-fifty probability value for $L(t)$) | $\text{median}=e^{M_L(t)}$ |
| Skewness | $\dfrac{K_3}{(K_2)^{\frac{3}{2}}}=\sqrt{e^{\sigma_L^2(t-ts)}-1}\left(e^{\sigma_L^2(t-ts)}+2\right)$ |
| Kurtosis | $\dfrac{K_4}{K_2^2}=e^{4\sigma_L^2(t-ts)}+2e^{3\sigma_L^2(t-ts)}+3e^{2\sigma_L^2(t-ts)}-6$ |

This is (with $ts=0$) just the 'mean value showing at the exponent' of the well-known ordinary (i.e. starting at $t=0$) GBM pdf, i.e.

$$\text{GBM}(t)_\text{pdf}\,(n;\,N_0,\,\mu_\text{GBM},\,\sigma_\text{GBM},\,t)=\frac{e^{-\frac{\left[\ln(n)-\left(\ln N_0+\left(\mu_\text{GBM}-\frac{\sigma_\text{GBM}^2}{2}\right)t\right)\right]^2}{2\sigma_\text{GBM}^2 t}}}{\sqrt{2\pi}\sigma_\text{GBM}\sqrt{tn}}\quad\text{with}\begin{cases}n>0,\\t\geqslant 0,\\N_0>0,\\\sigma_\text{GBM}\geqslant 0.\end{cases}\tag{12}$$

A summary of the statistical properties of the GBMs is given in Table 2. We conclude this short description of the GBM as the exponential sub-case of the general lognormal process (1) by *warning* that GBM is a misleading name, since GBM is a lognormal process and not Gaussian one, as the Brownian motion is indeed.

(2) Another particularly interesting case of the mean value function $m_L(t)$ in (3) is when it equals a generic *polynomial in t*, namely

$$m_\text{polynomial}(t)=\sum_{k=0}^{\text{degree}}c_k\,t^k,\tag{13}$$

$c_k$ being the coefficient of the $k$th power of the time $t$ in the polynomial (13). We just confine ourselves to mention

Table 2. *Summary of the properties of the lognormal distribution that applies to the GBM stochastic process N(t) as the exponentially increasing number of ET communicating civilizations in the Galaxy, as well as the number of living species on Earth over the last 3.5 billion years (Malthusian or exponential growth).*

| | |
|---|---|
| Stochastic process | $N(t) = \begin{cases} \text{(1) Number of ET civilizations (in SETI)} \\ \text{(2) Number of living species (in evolution)} \end{cases}$ |
| Probability distribution | Lognormal distribution of the geometric Brownian motion (GBM), i.e. the lognormal stochastic process with exponential mean |
| Probability density function | $\text{GBM}(t)_\text{pdf}\ (n; \sigma_L, ts, Ns, t) = \dfrac{e^{-\frac{\left[\ln(n) - \left(\ln Ns + \left(\mu_{\text{GBM}} - \frac{\sigma^2_{\text{GBM}}}{2}\right)(t-ts)\right)\right]^2}{2\sigma_L^2(t-ts)}}}{\sqrt{2\pi}\,\sigma_{\text{GBM}}\sqrt{t-ts}\,n}$ with $\begin{cases} n > 0 \\ t \geqslant ts \\ Ns > 0 \\ \sigma_{\text{GBM}} \geqslant 0 \end{cases}$ |
| Mean value | $\langle \text{GBM}(t) \rangle \equiv m_{\text{GBM}(t)} = Ns\, e^{\mu(t-ts)}$ |
| Variance | $\sigma^2_{\text{GBM}(t)} = Ns^2\, e^{2\mu(t-ts)}\left(e^{\sigma^2_{\text{GBM}}(t-ts)} - 1\right)$ |
| Standard deviation | $\sigma_{\text{GBM}(t)} = Ns\, e^{\mu(t-ts)}\sqrt{e^{\sigma^2_{\text{GBM}}(t-ts)} - 1}$ |
| Upper standard deviation curve | $m_{\text{GBM}(t)} + \sigma_{\text{GBM}(t)} = Ns\, e^{\mu(t-ts)}\left(1 + \sqrt{e^{\sigma^2_{\text{GBM}}(t-ts)} - 1}\right)$ |
| Lower standard deviation curve | $m_{\text{GBM}(t)} - \sigma_{\text{GBM}(t)} = Ns\, e^{\mu(t-ts)}\left(1 - \sqrt{e^{\sigma^2_{\text{GBM}}(t-ts)} - 1}\right)$ |
| All the moments, i.e. $k$th moment | $\langle \text{GBM}^k(t) \rangle = Ns^k\, e^{\left(k\mu - (k-k^2)\frac{\sigma^2_{\text{GBM}}}{2}\right)(t-ts)}$ |
| Mode (= abscissa of the lognormal peak) | $n_{\text{mode}} \equiv n_{\text{peak}} = Ns\, e^{\left(\mu - 3\frac{\sigma^2_{\text{GBM}}}{2}\right)(t-ts)}$ |
| Value of the mode peak | $f_{\text{GBM}(t)}(n_{\text{mode}}) = \dfrac{e^{\left(\sigma^2_{\text{GBM}} - \mu\right)(t-ts)}}{Ns\sqrt{2\pi}\,\sigma_{\text{GBM}}\sqrt{t-ts}}$ |
| Median (= fifty-fifty probability value for GBM($t$)) | $\text{median} = Ns\, e^{\left(\mu - \frac{\sigma^2_{\text{GBM}}}{2}\right)(t-ts)}$ |
| Skewness | $\dfrac{K_3}{(K_2)^{\frac{3}{2}}} = \sqrt{e^{\sigma^2_{\text{GBM}}(t-ts)} - 1}\left(e^{\sigma^2_{\text{GBM}}(t-ts)} + 2\right)$ |
| Kurtosis | $\dfrac{K_4}{K_2^2} = e^{4\sigma^2_{\text{GBM}}(t-ts)} + 2e^{3\sigma^2_{\text{GBM}}(t-ts)} + 3e^{2\sigma^2_{\text{GBM}}(t-ts)} - 6$ |

here that the case where (13) is a second-degree polynomial (i.e. a parabola in $t$) may be used to describe the mass extinctions that plagued life on Earth over the last 3.5 billion years, as we shall see in the section 'Mass extinctions described by an adjusted parabola branch' of this paper.

A summary of the statistical properties of the $L(t)$ process when its mean value is the polynomial (13) is given in Table 3.

## Introducing *b*-lognormals

We must also introduce the notion of $b$-lognormal pdf, namely a lognormal pdf (in the time variable as independent variable), which rather than starting at $t=0$, starts at any time $t=b$. Therefore, the $b$-lognormal pdf is given by

$$b\text{-lognormal_pdf}(t;\ \mu,\ \sigma,\ b) = \frac{e^{-\frac{[\ln(t-b)-\mu]^2}{2\sigma^2}}}{\sqrt{2\pi}\,\sigma\,(t-b)}. \tag{14}$$

It describes the lifetime of any living being, be it a cell, a plant, a human, a civilization of humans or even an ET civilization. Interested readers should please read Maccone (2013), particularly pp. 227–245 where the notion of *finite* (*in time*) *b-lognormal* (as opposed to the infinite (in time) $b$-lognormal given by (14)) was also introduced.

Professional statisticians sometime call 'three-parameter lognormal' the pdf (14). This is because (14) embodies the third parameter $b$ in addition to the two classical ones, $\mu$ and $\sigma$ of the ordinary lognormal, i.e. (14) with $b=0$. Statisticians are obviously interested in the numerical estimation of $\mu$ and $\sigma$, and also of $b$, by virtue of the 'maximum Likelihood' techniques of statistics. Although that is certainly an important topic for the application of $b$-lognormals to real cases, we are not going to face these issues in this paper: their study has to be delayed to a further research paper.

## Peak-Locus Theorem

The Peak-Locus theorem is a new mathematical discovery of ours which plays a central role in Evo-SETI theory. In its most general formulation, it holds good for any lognormal process $L(t)$ and any arbitrary function $M_L(t)$ (or mean value $m_L(t)$).

In words, and utilizing the simple example of the Peak-Locus theorem applied to GBMs, the Peak-Locus theorem states what is shown in Fig. 1: the family of all $b$-lognormals

Table 3. *Summary of the properties of the polynomial lognormal distribution that applies to the stochastic process P(t) as the lognormally changing number of ET communicating civilizations in the Galaxy, as well as the number of living species on Earth over the last 3.5 billion years, if the mean value is a polynomial in the time.*

| | |
|---|---|
| Stochastic process | $P(t) = \begin{cases} \text{(1) Number of ET civilizations (in SETI)} \\ \text{(2) Number of living species (in evolution)} \end{cases}$ |
| Probability distribution | Lognormal distribution of stochastic processes with polynomial mean |
| Probability density function | $P(t)_\text{pdf}\left(n;\ \sum_{k=0}^{\text{degree}} c_k(t-ts)^k,\ \sigma_{\text{polynomial}},\ t\right) = \dfrac{e^{-\frac{\left[\ln(n)-\log\left(\sum_{k=0}^{\text{degree}} c_k(t-ts)^k\right)+\frac{\sigma^2_{\text{polynomial}}}{2}(t-ts)\right]^2}{2\sigma^2_{\text{polynomial}}(t-ts)}}}{\sqrt{2\pi}\,\sigma_{\text{polynomial}}\sqrt{t-ts}\,n}$ with $\begin{cases} n > 0 \\ t \geqslant ts \\ \sigma_{\text{polynomial}} \geqslant 0 \end{cases}$ |
| Mean value curve | $\langle P(t)\rangle \equiv m_{P(t)} = \sum_{k=0}^{\text{degree}} c_k(t-ts)^k = e^{M_{\text{polynomial}}(t)} e^{\frac{\sigma^2_{\text{polynomial}}}{2}(t-ts)}$ |
| Variance | $\sigma^2_{P(t)} = \left(\sum_{k=0}^{\text{degree}} c_k(t-ts)^k\right)^2 \left(e^{\sigma^2_{\text{polynomial}}(t-ts)} - 1\right)$ |
| Standard deviation | $\sigma_{P(t)} = \left(\sum_{k=0}^{\text{degree}} c_k(t-ts)^k\right)\sqrt{e^{\sigma^2_{\text{polynomial}}(t-ts)} - 1}$ |
| Upper standard deviation curve | $m_{P(t)} + \sigma_{P(t)} = \left(\sum_{k=0}^{\text{degree}} c_k(t-ts)^k\right)\left(1 + \sqrt{e^{\sigma^2_{\text{polynomial}}(t-ts)} - 1}\right)$ |
| Lower standard deviation curve | $m_{P(t)} - \sigma_{P(t)} = \left(\sum_{k=0}^{\text{degree}} c_k(t-ts)^k\right)\left(1 - \sqrt{e^{\sigma^2_{\text{polynomial}}(t-ts)} - 1}\right)$ |
| All the moments, i.e. *j*th moment | $\langle P^j(t)\rangle = \left(\sum_{k=0}^{\text{degree}} c_k(t-ts)^k\right)^j e^{(j^2-j)\frac{\sigma^2_{\text{polynomial}}}{2}(t-ts)}$ |
| Mode (abscissa of the lognormal peak) | $n_{\text{mode}} \equiv n_{\text{peak}} = \left(\sum_{k=0}^{\text{degree}} c_k(t-ts)^k\right) e^{-3\frac{\sigma^2_{\text{polynomial}}}{2}(t-ts)}$ |
| Value of the mode peak | $f_{N(t)}(n_{\text{mode}}) = \dfrac{e^{\sigma^2_{\text{polynomial}}(t-ts)}}{\left(\sum_{k=0}^{\text{degree}} c_k(t-ts)^k\right)\sqrt{2\pi}\,\sigma_{\text{polynomial}}\sqrt{t-ts}}$ |
| Median (= fifty-fifty probability value for $P(t)$) | $\text{median} = \left(\sum_{k=0}^{\text{degree}} c_k(t-ts)^k\right) e^{-\frac{\sigma^2_{\text{GBM}}}{2}(t-ts)}$ |
| Skewness | $\dfrac{K_3}{(K_2)^{\frac{3}{2}}} = \sqrt{e^{\sigma^2_{\text{polynmial}}(t-ts)} - 1}\left(e^{\sigma^2_{\text{polynomial}}(t-ts)} + 2\right)$ |
| Kurtosis | $\dfrac{K_4}{K_2^2} = e^{4\sigma^2_{\text{polynomial}}(t-ts)} + 2e^{3\sigma^2_{\text{polynomial}}(t-ts)} + 3e^{2\sigma^2_{\text{polynomial}}(t-ts)} - 6$ |

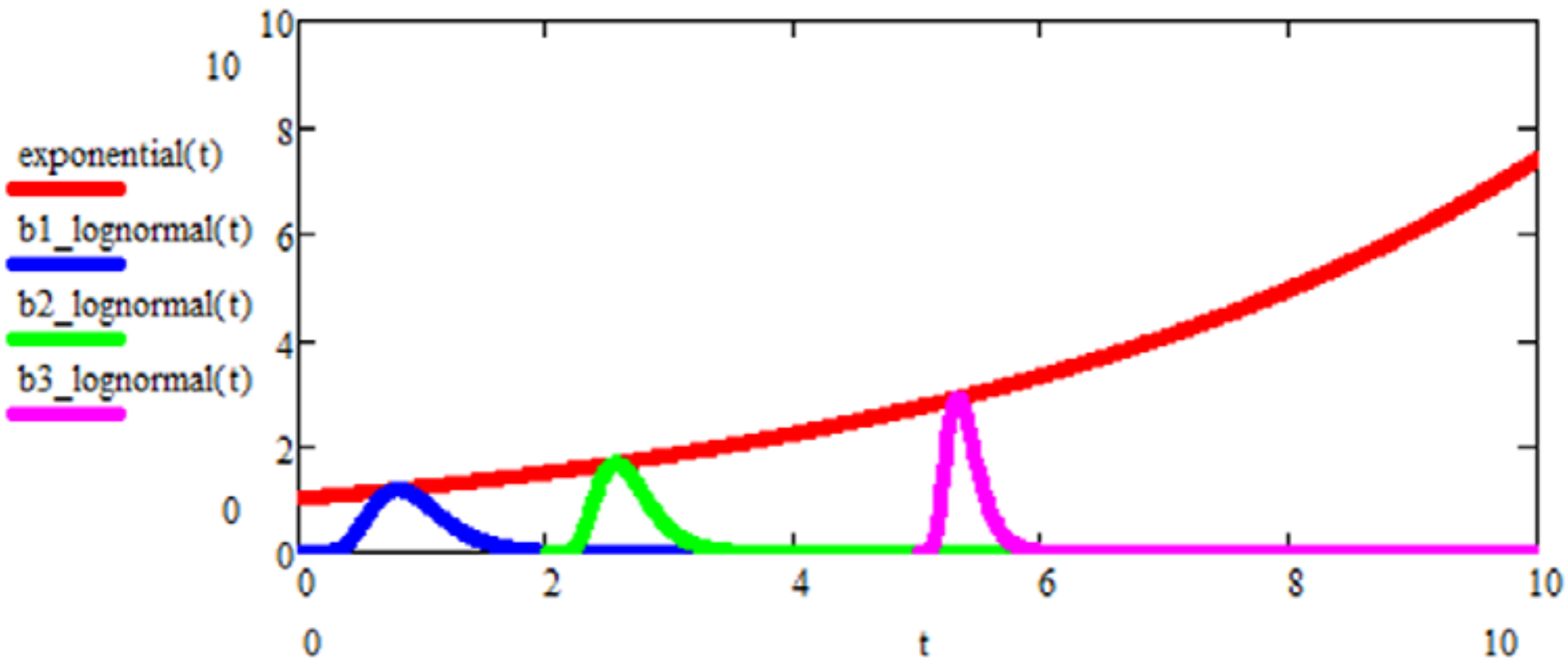

**Fig. 1.** Darwinian exponential as the geometric *locus of the peaks* of *b*-lognormals. Each *b*-lognormal is a lognormal starting at a time ($t = b =$ birth time) in general different from zero and represents a different *species* that originated at time *b* of the Darwinian Evolution. That is Cladistics in our Evo-SETI model. It is evident that, the more the generic 'running *b*-lognormal' moves to the right, its peak becomes higher and higher and narrower and narrower, since the area under the *b*-lognormal always equals 1. Then, the (Shannon) *entropy* of the running *b*-lognormal is the *degree of evolution* reached by the corresponding *species* (or living being, or civilization, or ET civilization) in the course of evolution.

'trapped' between the time axis and the growing exponential of the GBMs, where all the $b$-lognormal peaks lie, can be *exactly* (i.e. without any numerical approximation) described by three equations yielding the three parameters μ, σ and $b$ as three functions of the peak abscissa, $p$, only.

In equations, the Peak-Locus theorem states that the family of $b$-lognormals having each of its peak exactly located *on* the mean value curve (4), is given by the following three equations, specifying the parameters $\mu(p)$, $\sigma(p)$ and $b(p)$, appearing in (14) as three functions of the single 'independent variable' $p$, i.e. the abscissa (i.e. the time) of the $b$-lognormal's peak:

$$\begin{cases} \mu(p) = \dfrac{e^{\sigma_L^2 ts} e^{-2[M_L(p)-M_L(ts)]}}{4\pi N s^2} - p\dfrac{\sigma_L^2}{2}, \\ \sigma(p) = \dfrac{e^{\frac{\sigma_L^2 ts}{2}} e^{-[M_L(p)-M_L(ts)]}}{\sqrt{2\pi} N s}, \\ b(p) = p - e^{\mu(p)-[\sigma(p)]^2}. \end{cases} \quad (15)$$

This general form of the Peak-Locus theorem is proven in the Appendix available in the supplementary material by equations (%i28) through (%o44). The remarkable point about all this seems to be the exact separability of all the equations involved in the derivation of (15), a fact that was unexpected to this author when he discovered it around December 2013. And the consequences of this new result are in the applications:

(1) For instance in the 'parabola model' for mass extinctions that will be studied in the section 'Mass extinctions described by an adjusted parabola branch' of this paper.
(2) For instance the Markov–Korotayev cubic that will be studied in the section 'Markov–Korotayev biodiversity regarded as a lognormal stochastic process having a cubic mean value' of this paper.
(3) And finally in the many stochastic processes each having a cubic mean value that are just the natural extension into statistics of the deterministic cubics studied by this author in Chapter 10 of his book 'Mathematical SETI', (Maccone 2012). But the study of the entropy of all these cubic lognormal processes has to be deferred to a future research paper.

Notice now that, in the particular case of the GBMs having mean value (9) with $\mu_{GBM} = B$, and starting at $ts = 0$ with $N_0 = Ns = A$, the Peak-Locus theorem (15) boils down to the simpler set of equations

$$\begin{cases} \mu(p) = \dfrac{1}{4\pi A^2} - Bp, \\ \sigma = \dfrac{1}{\sqrt{2\pi} A}, \\ b(p) = p - e^{\mu(p)-\sigma^2}. \end{cases} \quad (16)$$

In this simpler form, the Peak-Locus theorem was already published by the author in Maccone (2011b, 2012, 2013), while its most general form (15) is new for this paper.

## Entropy as the evolution measure

The (Shannon) entropy of the running $b$-lognormal

$$H_L(p) = \frac{1}{\ln(2)}\left[\ln(\sqrt{2\pi}\,\sigma(p)) + \mu(p) + \frac{1}{2}\right] \quad (17)$$

is a function of the peak abscissa $p$ and is measured in bits, as common in Shannon's information theory. By virtue of the Peak-Locus theorem (15), it becomes (see the Appendix available in the supplementary material, (%o45) through (%o50))

$$H_L(p) = \frac{\dfrac{e^{\sigma_L^2 ts - 2[M_L(p)-M_L(ts)]}}{4\pi N s^2} - \dfrac{\sigma_L^2(p-ts)+2[M_L(p)-M_L(ts)]}{2} + \dfrac{1-2\ln(Ns)}{2}}{\ln(2)}. \quad (18)$$

More precisely, (18) is the entropy of the family of $\infty^1$ running $b$-lognormals (the family's parameter is $p$) that are peaked *upon* the mean value curve (3). Although (3) is *not* the 'envelope' of the $b$-lognormals (14) in a strict mathematical sense, yet, in the practice, it is approximately the same thing, since it 'almost envelopes' all the $b$-lognormals.

This is 'the greatest result' of our Evo-SETI model inasmuch as, for instance, in the case of the history of the Western civilizations since the Greeks up to 2200 A.D. (represented each by a $b$-lognormal), as shown in Fig. 2, then the 'enveloping exponential' is just the GBM mean value exponential (see Maccone (2012, 2013) for more historic and mathematical details). So it also happens for Darwinian Evolution (see Maccone (2011b)).

The $b$-lognormal entropy (17) is thus the *measure of evolution amount* of that $b$-lognormal: it measures 'the decreasing disorganization in time of what that $b$-lognormal represents', let it be a cell, a plant, a human or a civilization of humans (since 'the product of many $b$-lognormals is one more $b$-lognormal), or even of ETs.

Entropy is thus disorganization decreasing in time. But would it not be more intuitive to use a measure of 'increasing organization' in time? Of course yes. Our

$$\text{Evo_Index}_L(p) = -[H_L(p) - H_L(ts)] \quad (19)$$

Evo-Index is a function of $p$ that, however, has a minus sign in front, thus changing the decreasing trend of the (Shannon) entropy (17) into the increasing trend of our Evo-Index (19). Also, our Evo-Index starts at zero at the initial time $ts$:

$$\text{Evo_Index}_L(ts) = 0. \quad (20)$$

Please see the Appendix (available in the supplementary material), (%i52) through (%o56).

In the GBM case, i.e. when the mean value (3) is the (Malthusian) exponential curve (9), the $b$-lognormal entropy (18) becomes just a *linear function of time p*,

$$H_{GBM}(p) = \frac{-\ln(A) + \dfrac{1}{4\pi A^2} - pB + \dfrac{1}{2}}{\ln(2)}. \quad (21)$$

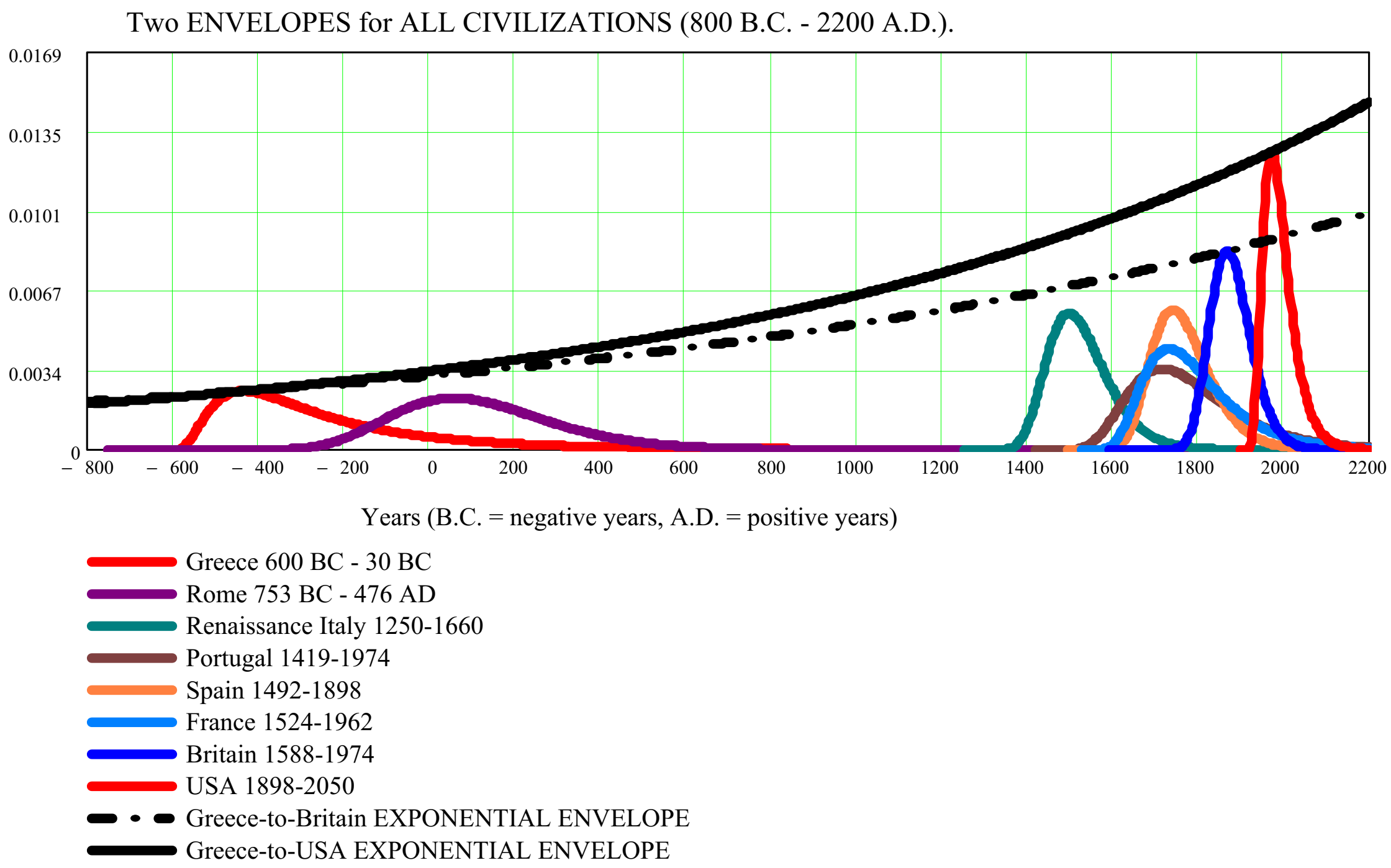

**Fig. 2.** Shown here are the eight leading civilizations of the Western world in the historic time-span between 800 B.C. and 2200 A.D. Each one is represented by a different $b$-lognormal given by an equation (14), and the 'envelope' of all of them is approximately given by the GBM exponential (9) (*Peak-Locus theorem*). Then, the (Shannon) *entropy* (17) of each $b$-lognormal becomes the *measure of the degree of evolution* reached by each civilization. Please see Maccone (2013, pp. 233–239) for more detailed descriptions and calculations.

Then, the Evo-Index of GBM simply is a new *linear function of time p also*

$$\text{Evo_Index}_{\text{GBM}}(p) = \frac{B}{\ln(2)}(p - ts). \tag{22}$$

That is, of course, a straight line starting at the time $ts$ of the origin of life on Earth and *increasing linearly thereafter, since it has the positive constant derivative*

$$\frac{\text{dEvo_Index}_{\text{GBM}}(p)}{\text{d}p} = \frac{B}{\ln(2)} = \text{a positive constant.} \tag{23}$$

But this is precisely the linear growth in time of the molecular clock also!

So, we have discovered that the entropy of our Evo-SETI model and the molecular clock are the same thing, apart for multiplicative constants (depending on the adopted units, such as bits, seconds, etc).

This is a great conclusion proving that our GBM model for Darwinian Evolution, described in Maccone (2011b, 2012, 2013) is correct.

We are 'grateful' to Emil Zuckerkandl, Linus Pauling, Motoo Kimura and his pupils Tomoko Ohta and Takeo Maruyama ('Neutral Theory of Molecular Evolution') for bringing the molecular clock to light (Nei & Sudhir 2000; Nei 2013).

Incidentally, the result that our entropy agrees with the molecular clock (apart from constants) is published in this paper for the first time.

## Evo-SETI

Every day astronomers are discovering new extra-solar planets, either by observations from the Earth or by space missions, like 'CoRoT' and 'Kepler'. 'Gaia' is now on its way to the Lagrangian point L2 of the Sun–Earth system, and will measure the parallaxes (= distances) of a billion stars in our Galaxy. A recent estimate sets at 40 billion the number of Earth-sized planets orbiting in the habitable zones of Sun-like stars and red dwarf stars within the Milky Way. Thus the assumption that 'we are alone in the Galaxy' (let alone the Universe), is becoming simply more and more foolish.

SETI is a branch of science trying to detect signatures of intelligent life: either by picking up a radio signal or by detecting an optical pulsating laser, or even by detecting a pulsating beam of neutrinos. SETI scientists are usually elected members of the SETI Permanent Committee of the International Academy of Astronautics (IAA) and, on 3 October 2012, this author was elected Chair of that Committee. So, the author now bears the responsibility to coordinate the world-wide SETI activities, and in this position, formulated the mathematical model of Evo-SETI summarized in this paper. Why?

The model was developed because we only have one example of a civilization (ourselves) that evolved to the point of developing technological capabilities like those required by SETI (radio and optical top instrumentation, super-computers, etc.). The Drake equation (1961) was the first step in theoretical SETI, and in Maccone (2008) the author gave its mathematical extension into probability and statistics.

But this is not enough. Suppose that one day the SETI scientists detect an alien message or a proof that Aliens exist, perhaps not too far from us in the Galaxy. What would we do then?

As Chair of the IAA SETI Permanent Committee, this author would firstly ask: how much *more advanced than us* are those ‘guys’? Especially in technology, if they were able to send signals, or other artefacts capable of being discovered by us. Thus, we need some scientific criterion capable of letting us know the technology gap between us and them, even if only approximately. This author thinks that the answer to this question is the entropy of the running $b$-lognormal, as described earlier. For instance, in Maccone (2013), the author estimated that the technology gap between Aztecs and Spaniards, when they suddenly met in 1519, was about fifty centuries, corresponding to an *entropy gap of 3.84 bits per individual.* It was this gap that made 20 million Aztecs to have a psychological breakdown and collapse in front of a few thousand ‘much superior’ Spaniards. We must think of that if we want to prepare for the first contact with an alien civilization.

## Mass extinctions of Darwinian Evolution described by a decreasing GBM

*GBMs to understand mass extinctions of the past*

In this section, we describe the use of GBMs to model the mass extinctions that occurred on Earth several times in its geological past. The most notable example probably is the mass extinction of dinosaurs 64 million years ago, now widely recognized by scientists as caused by the impact of a $\sim$10 km sized asteroid where the Chicxulub crater in Yucatan, Mexico, is now found (Alvarez *et al.* 1980; Alvarez 2008). Incidentally, in 2007 this author was part of a NASA team in charge of studying a space mission capable of deflecting an asteroid off its collision course against the Earth, should this event unfortunately occur again in the future: so he got a background in planetary defence.

Let us now go straight to the GBMs and consider the mean value given in the fourth line of Table 2 again, that is the mean value of a GBM *increasing* in time to simulate the rise of more and more species in the course of evolution, so $\mu > 0$ for it. But in modelling mass extinctions, we clearly must have a *decreasing* GBM, i.e. $\mu < 0$, over a much shorter time lapse, just years or some centuries instead of billions of years, as in Darwinian Evolution. So, the starting time now is the impact time, $ts = t_{\text{Impact}}$, and our GMB mean value becomes

$$\text{mean_value}(t) = C\, e^{\mu(t - t_{\text{Impact}})}, \tag{24}$$

where $C$ is a constant that we now determine. Just think that, at the impact time, (24) yields

$$\text{mean_value}(t_{\text{Impact}}) = C. \tag{25}$$

On the other hand, at the same impact time, one has

$$\text{mean_value}(t_{\text{Impact}}) = N_{\text{Impact}}, \tag{26}$$

where $N_{\text{Impact}}$ is the number of living species on Earth just seconds before the asteroid impact time. Thus, (25) and (26) immediately yield

$$C = N_{\text{Impact}}. \tag{27}$$

This, inserted into (24), yields the final mean value curve as a function of the time

$$\text{mean_value}(t) = N_{\text{Impact}} e^{\mu(t - t_{\text{Impact}})}. \tag{28}$$

Let us now consider what happens after the impact, namely the death of many living species over a period of time called ‘nuclear winter’ and caused by the debris thrown into the Earth’s atmosphere by the asteroid ejecta. Nobody seems to know exactly how long the nuclear winter lasted after the impact that actually killed all dinosaurs and other species, but not the mammals, who, being much smaller and so much more easy-fed, could survive the nuclear winter. Mathematically, let us call $t = t_{\text{End}}$ the time when the nuclear winter ended, so that the overall time span of the mass extinction is given by

$$t_{\text{End}} - t_{\text{Impact}}. \tag{29}$$

At time $t_{\text{End}}$, a certain number of living species, say $N_{\text{End}}$, survived. Replacing this into (28) yields

$$N_{\text{End}} = \text{mean_value}(t_{\text{End}}) = N_{\text{Impact}} e^{\mu(t_{\text{End}} - t_{\text{Impact}})}. \tag{30}$$

Solving (30) for $\mu$ yields *the first basic formula for our GBM model of mass extinctions*:

$$\mu = -\frac{\ln\left(\dfrac{N_{\text{Impact}}}{N_{\text{End}}}\right)}{t_{\text{End}} - t_{\text{Impact}}}. \tag{31}$$

Notice that in (31) are four input variables

$$(t_{\text{Impact}},\quad N_{\text{Impact}},\quad t_{\text{End}},\quad N_{\text{End}}), \tag{32}$$

which we must assign numerically in order determine $\mu$ for that particular mass extinction.

Let us also remark that it is convenient to introduce two new variables, time_lapse and $t_{\text{Extinction}}$, respectively, defined as the overall amount of time during which the extinction occurs, and the middle instance in this overall time lapse, namely

$$\begin{cases} \text{Time_Lapse} = t_{\text{End}} - t_{\text{Impact}}, \\ t_{\text{Extinction}} = \dfrac{t_{\text{Impact}} + t_{\text{End}}}{2}. \end{cases} \tag{33}$$

Clearly (31), by virtue of (33), becomes

$$\mu = -\frac{\ln\left(\dfrac{N_{\text{Impact}}}{N_{\text{End}}}\right)}{\text{Time_Lapse}}. \tag{34}$$

This version of (31) is easier to differentiate as it only has three independent variables instead of four. Thus, the total differential of (34) is found (but we will not write all the steps here), and, once divided by (34), yields the relative error on $\mu$ expressed in terms of the relative errors on $N_{\text{Impact}}$, $N_{\text{End}}$

and Time_Lapse:

$$\frac{\delta\mu}{\mu} = -\frac{\delta\text{Time_Lapse}}{\text{Time_Lapse}} + \frac{1}{\ln\left(\frac{N_{\text{Impact}}}{N_{\text{End}}}\right)}\frac{\delta N_{\text{Impact}}}{N_{\text{Impact}}} - \frac{1}{\ln\left(\frac{N_{\text{Impact}}}{N_{\text{End}}}\right)}\frac{\delta N_{\text{End}}}{N_{\text{End}}}. \tag{35}$$

Let us now find $\sigma$.

To this end, we must introduce a fifth input (besides the four input variables given by (32)), denoted $\delta N_{\text{End}}$ and representing the standard deviation affecting the number of living species on Earth at the end of the nuclear winter, i.e. when life starts growing again. This means that we must now consider the GBM as a standard deviation function of the time, $\Delta(t)$, given by the sixth line in Table 2, which in this case, takes the form

$$\Delta(t) = N_{\text{Impact}} e^{\mu(t-t_{\text{Impact}})}\sqrt{e^{\sigma^2(t-t_{\text{Impact}})} - 1}. \tag{36}$$

At the end time, $t_{\text{End}}$, (36) becomes

$$\Delta(t_{\text{End}}) = N_{\text{Impact}} e^{\mu(t_{\text{End}}-t_{\text{Impact}})}\sqrt{e^{\sigma^2(t_{\text{End}}-t_{\text{Impact}})} - 1}. \tag{37}$$

But this equals $\delta N_{\text{End}}$ by the very definition of $\delta N_{\text{End}}$, and so we get the new equation

$$\delta N_{\text{End}} = N_{\text{Impact}} e^{\mu(t_{\text{End}}-t_{\text{Impact}})}\sqrt{e^{\sigma^2(t_{\text{End}}-t_{\text{Impact}})} - 1}. \tag{38}$$

This is basically the equation in $\sigma$ we were seeking. We only have to replace $\mu$ into (38) by virtue of (34), and then solve the resulting equation for $\sigma$. By doing so (we omit the relevant steps for the sake of brevity), we finally get the sought expression of $\sigma$:

$$\sigma = \sqrt{\frac{\ln\left[1 + \left(\frac{\delta N_{\text{End}}}{N_{\text{End}}}\right)^2\right]}{\text{Time_Lapse}}}. \tag{39}$$

This is the GBM $\sigma$ for the mass extinctions.

Notice that the special $\delta N_{\text{END}} = 0$ case of (39), immediately yields $\sigma = 0$. This is the special case where the GBM 'converges' (so as to say) into a *single point* at $t = t_{\text{END}}$, namely, with probability one there will be *exactly* $N_{\text{End}}$ species that survived the nuclear winter after the impact. This is just like the initial condition of ordinary Brownian motion, $B(0) = 0$, which is always fulfilled with probability one. But in this case it is a *final condition*, rather than an initial condition. As such, this particular case of (39) is hardly realistic in the true world of an after-impact. Nevertheless, we wanted to point it out just to show how subtle the mathematics of stochastic processes can be.

Another remark following from (39) is about the expression of the relative error on $\sigma$, namely $\delta\sigma/\sigma$, expressed in terms of the four inputs (32) plus $\delta N_{\text{End}}$. The relevant expression is long and complicated, and we will not rewrite it here.

Finally, it must to be mentioned that the upper standard deviation curve given by (28) plus (36), i.e.

$$\text{upper_st_dev}(t) = N_{\text{Impact}} e^{\mu(t-t_{\text{Impact}})}\left[1 + \sqrt{e^{\sigma^2(t-t_{\text{Impact}})} - 1}\right] \tag{40}$$

has its maximum at the just-after-impact time

$$t_{\text{Impact}} + \frac{1}{\sigma^2}\ln\left[\frac{2\mu\left(\sqrt{\mu^2 - 2\mu\sigma^2 - \sigma^4}\right) + \sigma^2 + 3\mu}{(\sigma^2 + 2\mu)^2}\right]. \tag{41}$$

Again, we will not rewrite here all the steps leading to (41), and just confine ourselves to mentioning that one gets a quadratic in $e^{\sigma^2 t}$ that, solved for $t$, yields (41).

Having given the mathematical theory of mass extinctions provided by GBMs, we now proceed to show a numerical example. Naturally, the chosen example is about the Cretaceous–Paleogene (K–Pg) impact and the ensuing nuclear winter, which we assume to have lasted a thousand years after the impact itself, though other shorter time lapses could be considered as well.

### *Example: the K–Pg mass extinction extending 10 centuries after impact*

Readers should be warned that the numeric example and graph we now present is just an exercise, and we do not claim that it really shows what happened 64 million years ago during the K–Pg impact and the consequent mass extinction. Yet it provides useful hints about how the GBMs work in the simulations of true mass extinctions, and not just those of the past, but also those of the future, should an asteroid hit the Earth again and cause millions or billions of human casualties: planetary defence is a 'must' for us!

So, let us assume that:

(1) The K–Pg impact occurred *exactly* 64 million years ago (just to simplify the calculations):

$$t_{\text{Impact}} = -64 \times 10^6 \text{ year}. \tag{42}$$

(2) At impact, there were 100 living species on Earth,

$$N_{\text{Impact}} = 100. \tag{43}$$

Again, this is likely to be very roughly underestimated, but we use 100 so as to immediately draw the percentage of surviving species as described at the next point (3).

(3) At the end of the impact effects, there were only 30 living species, and only the 30% survived,

$$N_{\text{End}} = 30. \tag{44}$$

(4) We also assume that the error on the value of (44) is about 33.3%. In other words, we assume

$$\delta N_{\text{End}} = 10. \tag{45}$$

(5) Finally, we assume that the impact effects lasted for a 1000 years, i.e.

$$t_{\text{End}} - t_{\text{Impact}} = 1000 \text{ year}, \tag{46}$$

from which, by virtue of (42), we infer

$$t_{\text{End}} = -63.999 \times 10^6 \text{ year}. \tag{47}$$

These are our five input data. The two outputs then are

$$\mu = -3.815 \times 10^{-11}\ \text{s}^{-1} = -1.204 \times 10^{-3}\ \text{year}^{-1} \tag{48}$$

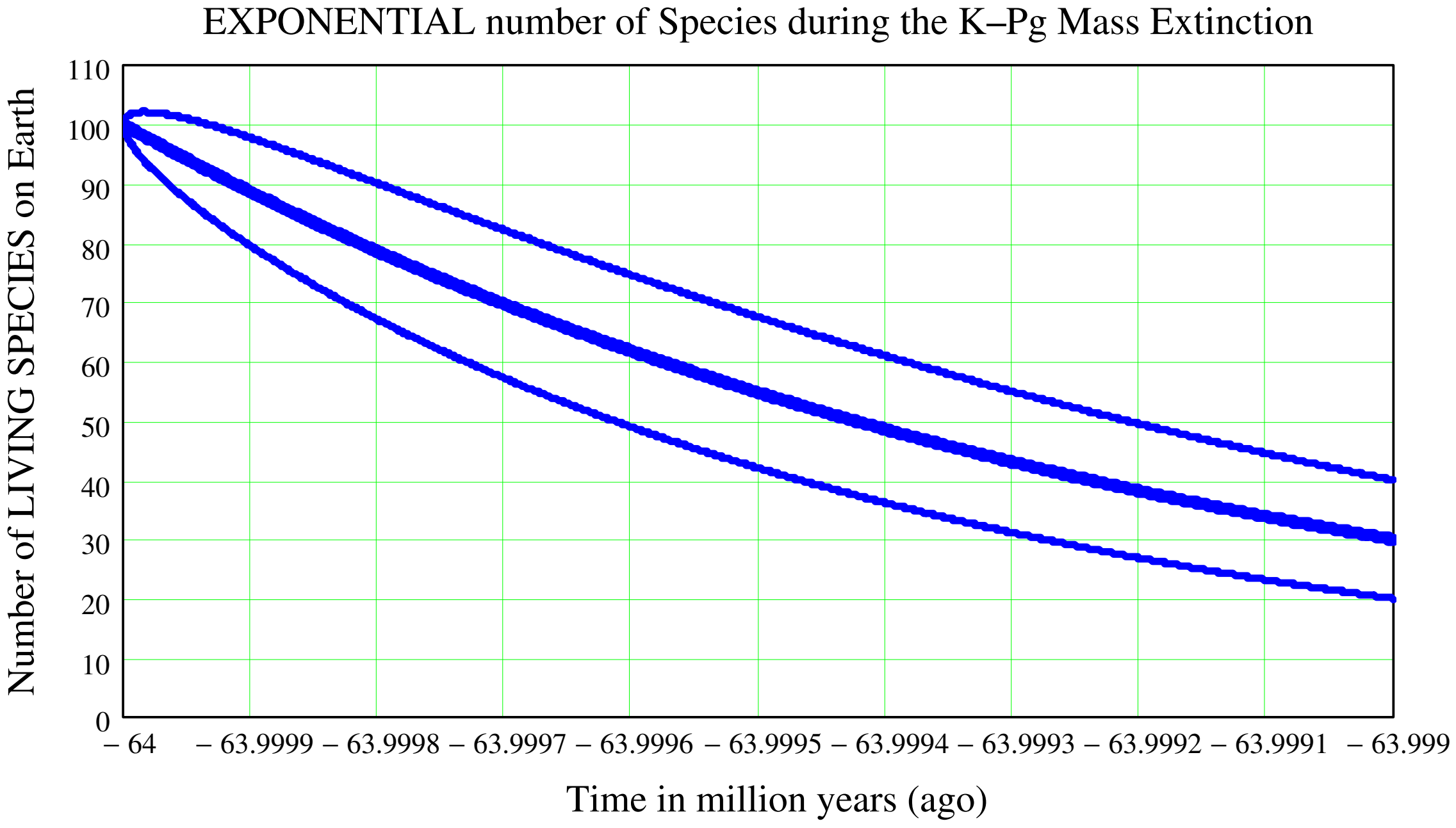

**Fig. 3.** The K–Pg mass extinction as a decreasing GBM in the number of living species over 1000 years after impact. The maximum of the upper standard deviation curve has the numeric value $-6.3999983 \times 10^7$ years given by (41).

and

$$\sigma^2 = 3.339 \times 10^{-12}\,\text{s}^{-1} = 1.054 \times 10^{-4}\,\text{year}^{-1}. \tag{49}$$

Figure 3 shows the mean value curve (solid blue curve), and the two upper and lower standard deviation curves (solid thin blue curves) for the corresponding GBM in the decreasing number of living species on Earth as the consequence of the impact.

In conclusion, Table 4 summarizes all results about the decreasing GBM representing the decreasing number of species on Earth during a mass extinction.

In the future, these ideas should be extended not just to the analysis of all mass extinctions that occurred in the geological past of Earth, but also to crucial events in human history such as wars, famines, epidemics and so on, when mass extinctions of humans occurred. An excellent topic to describe mathematically large sections of history that so far were mostly described by means of words only. By doing so, we would significantly contribute to the studies on mathematical history.

## Mass extinctions described by an adjusted parabola branch

### *Adjusting the parabola to the mass extinctions of the past*

The mass extinction model described in the previous section and based on an adjusted and decreasing GBM has a flaw: the tangent straight line to its mean value curve at the end time is *not* horizontal. Thus, it is not a realistic model inasmuch as its end time cannot correctly represent the starting point after which the number of living species on Earth started growing once again.

On the contrary, the mass extinction model that we built in this section does not have any such flaw: its end time is both the end time of the decreasing number of living species on Earth and also its starting time for increasing living species numbers. Its tangent straight line is indeed horizontal, as required.

Getting now over to the mathematics, consider the adjusted mean value curve of $L(t)$ given by the parabola (i.e. second-order polynomial in the adjusted time $(t - t_{\text{Impact}})$

$$m_{\text{parabola}}(t) = c_2(t - t_{\text{Impact}})^2 + c_1(t - t_{\text{Impact}}) + c_0. \tag{50}$$

In order to find its three unknown coefficients $c_0$, $c_1$ and $c_2$, we must resort to the initial and final conditions (i.e. the two boundary conditions of the problem):

$$\begin{cases} m_{\text{parabola}}(t_{\text{Impact}}) = N_{\text{Impact}}, \\ m_{\text{parabola}}(t_{\text{End}}) = N_{\text{End}}. \end{cases} \tag{51}$$

Inserting (50) into (51), the latter takes the form

$$\begin{cases} N_{\text{Impact}} = m_{\text{parabola}}(t_{\text{Impact}}) = c_0, \\ N_{\text{End}} = c_2(t_{\text{End}} - t_{\text{Impact}})^2 + c_1(t_{\text{End}} - t_{\text{Impact}}) + c_0. \end{cases} \tag{52}$$

The last two equations reduce to the single one

$$N_{\text{End}} - N_{\text{Impact}} = c_2(t_{\text{End}} - t_{\text{Impact}})^2 + c_1(t_{\text{End}} - t_{\text{Impact}}). \tag{53}$$

On the other hand, the time derivative of the mean value (50) is

$$\frac{\mathrm{d}m_{\text{parabola}}(t)}{\mathrm{d}t} = 2c_2(t - t_{\text{Impact}}) + c_1. \tag{54}$$

Equating this to zero, and replacing the time by the end time, we impose that the tangent straight line at the end time must be horizontal. Thus, from (54) one gets:

$$2c_2(t_{\text{End}} - t_{\text{Impact}}) + c_1 = 0, \tag{55}$$

Table 4. *Summary of the properties of the lognormal distribution that applies to the stochastic process $N_{\mathrm{DEC}}(t)$ as the exponentially decreasing number of living species on Earth during a mass extinction.*

| | |
|---|---|
| Stochastic process | $N_{\mathrm{DECREASING}}(t) \equiv N_{\mathrm{DEC}}(t) =$ Number of living species (in a mass extinction) |
| Probability distribution | Lognormal distribution of the adjusted and decreasing GBM starting at $t_{\mathrm{Impact}}$ |
| Probability density function | $N_{\mathrm{DEC}}(t)_\mathrm{pdf}(n;\ \mu,\ \sigma,\ t_{\mathrm{Impact}},\ t) = \dfrac{e^{-\frac{\left[\ln(n)-\left(\ln(N_{\mathrm{Impact}})+\mu(t-t_{\mathrm{Impact}})-\frac{\sigma^2(t-t_{\mathrm{Impact}})}{2}\right)\right]^2}{2\sigma^2(t-t_{\mathrm{Impact}})}}}{\sqrt{2\pi}\,\sigma\,\sqrt{t-t_{\mathrm{Impact}}}\,n}$ with $\begin{cases} n > 0 \\ t \geqslant t_{\mathrm{Impact}} \\ N_{\mathrm{Impact}} > 0 \\ \sigma \geqslant 0 \end{cases}$ |
| Particular $M_{\mathrm{GBM}}(t)$ function | $M_{\mathrm{DEC}}(t) = \ln N_{\mathrm{Impact}} + \left(\mu - \frac{\sigma^2}{2}\right)(t - t_{\mathrm{Impact}})$ |
| Mean value curve | $\langle N_{\mathrm{DEC}}(t)\rangle \equiv m_{\mathrm{DEC}}(t) = N_{\mathrm{Impact}} e^{\mu(t-t_{\mathrm{Impact}})}$ |
| Variance | $\sigma^2_{\mathrm{DEC}(t)} = N^2_{\mathrm{Impact}} e^{2\mu(t-t_{\mathrm{Impact}})}(e^{\sigma^2(t-t_{\mathrm{Impact}})} - 1)$ |
| Standard deviation | $\sigma_{N_{\mathrm{DEC}}(t)} = N_{\mathrm{Impact}} e^{\mu(t-t_{\mathrm{Impact}})}\sqrt{e^{\sigma^2(t-t_{\mathrm{Impact}})} - 1}$ |
| Upper standard deviation curve | $m_{N_{\mathrm{DEC}}(t)} + \sigma_{N_{\mathrm{DEC}}(t)} = N_{\mathrm{Impact}} e^{\mu(t-t_{\mathrm{Impact}})}\left[1 + \sqrt{e^{\sigma^2(t-t_{\mathrm{Impact}})} - 1}\right]$ |
| Lower standard deviation curve | $m_{N_{\mathrm{DEC}}(t)} - \sigma_{N_{\mathrm{DEC}}(t)} = N_{\mathrm{Impact}} e^{\mu(t-t_{\mathrm{Impact}})}\left[1 - \sqrt{e^{\sigma^2(t-t_{\mathrm{Impact}})} - 1}\right]$ |
| All the moments, i.e. $k$th moment | $\langle N^k_{\mathrm{DEC}}(t)\rangle = N^k_{\mathrm{Impact}} e^{k\mu(t-t_{\mathrm{Impact}})} e^{(k^2-k)\frac{\sigma^2(t-t_{\mathrm{Impact}})}{2}}$ |
| Mode (=abscissa of the lognormal peak) | $n_{\mathrm{mode}} \equiv n_{\mathrm{peak}} = N_{\mathrm{Impact}} e^{\mu(t-t_{\mathrm{Impact}})} e^{-\frac{3\sigma^2(t-t_{\mathrm{Impact}})}{2}}$ |
| Value of the mode peak | $f_{N_{\mathrm{DEC}}(t)}(n_{\mathrm{mode}}) = \dfrac{1}{N_{\mathrm{Impact}}\sqrt{2\pi}\,\sigma\,\sqrt{t - t_{\mathrm{Impact}}}} e^{-\mu(t-t_{\mathrm{Impact}})} e^{\sigma^2(t-t_{\mathrm{Impact}})}$ |
| Median (=fifty-fifty probability value for $N(t)$) | $\mathrm{median} = m = N_{\mathrm{Impact}} e^{\mu(t-t_{\mathrm{Impact}})} e^{-\frac{\sigma^2(t-t_{\mathrm{Impact}})}{2}}$ |
| Skewness | $\dfrac{K_3}{(K_2)^{\frac{3}{2}}} = \sqrt{e^{\sigma^2_{\mathrm{polynomial}}(t-ts)} - 1}\left(e^{\sigma^2_{\mathrm{polynomial}}(t-ts)} + 2\right)$ |
| Kurtosis | $\dfrac{K_4}{(K_2)^2} = e^{4\sigma^2(t-t_{\mathrm{Impact}})} + 2e^{3\sigma^2(t-t_{\mathrm{Impact}})} + 3e^{2\sigma^2(t-t_{\mathrm{Impact}})} - 6$ |

which, solved for $c_1$ and matched to (53), yields

$$\begin{cases} N_{\mathrm{End}} - N_{\mathrm{Impact}} = -c_2(t_{\mathrm{End}} - t_{\mathrm{Impact}})^2, \\ c_1 = -2c_2(t_{\mathrm{End}} - t_{\mathrm{Impact}}). \end{cases} \tag{56}$$

These two linear equations in $c_1$ and $c_2$ may immediately be solved for them, with the result

$$\begin{cases} c_2 = \dfrac{N_{\mathrm{Impact}} - N_{\mathrm{End}}}{(t_{\mathrm{End}} - t_{\mathrm{Impact}})^2}, \\ c_1 = \dfrac{-2(N_{\mathrm{Impact}} - N_{\mathrm{End}})}{t_{\mathrm{End}} - t_{\mathrm{Impact}}}. \end{cases} \tag{57}$$

Finally, inserting both (57) and the upper equation (52) into the mean value parabola (50), the latter takes its final form

$$m_{\mathrm{parabola}}(t) = (N_{\mathrm{Impact}} - N_{\mathrm{End}})\left[\frac{(t - t_{\mathrm{Impact}})^2}{(t_{\mathrm{End}} - t_{\mathrm{Impact}})^2} - 2\frac{t - t_{\mathrm{Impact}}}{t_{\mathrm{End}} - t_{\mathrm{Impact}}}\right] + N_{\mathrm{Impact}}. \tag{58}$$

One may immediately check that the two boundary conditions (51) are indeed fulfilled by (58). Also, the minimum of the parabola (58) (i.e. the zero of its first time derivative) falls at the end time $t_{\mathrm{end}}$, obviously by construction, i.e. because of (54). So, the parabola (58) is indeed the right curve with a horizontal tangent line at the end, which we were seeking.

As for the standard deviation, it is given by the seventh row in Table 3, of course 'adjusted' by replacing the time $t$ appearing in Table 3 by the new time difference $(t - t_{\mathrm{Impact}})$ appearing in the mean value curve (58) already. Thus, the standard deviation for the parabolic mass extinction model is given by

$$\sigma_{\text{parabola}}(t) = m_{\text{parabola}}(t)\sqrt{e^{\sigma^2(t - t_{\text{Impact}})} - 1}. \quad (59)$$

Table 5. *Summary of the properties of the lognormal distribution that applies to the stochastic process $P_{\text{parabola}}(t)$ = decreasing number of living species on Earth during a mass extinction whose mean value decreases like the left-branch of a parabola between $t_{\text{Impact}}$ and $t_{\text{End}}$ (the parabola minimum, thus having a horizontal line tangent at $t = t_{\text{End}}$)*

| | |
|---|---|
| Stochastic process | $P_{\text{parabola}}(t)$ = Number of living species (in a parabolic mass extinction) |
| Probability distribution | Lognormal distribution of the adjusted and parabolic process starting at $t_{\text{Impact}}$ |
| Probability density function | $P_{\text{parabola}}(t)_\text{pdf}(n;\ M_{\text{parabola}}(t),\ \sigma,\ t) = \frac{1}{\sqrt{2\pi}\sigma\sqrt{tn}} e^{-\frac{[\ln(n) - M_{\text{parabola}}(t)]^2}{2\sigma^2 t}}$ for $n > 0$ |
| Particular $M_{\text{parabola}}(t)$ function | $M_{\text{parabola}}(t) = \ln\left((N_{\text{Impact}} - N_{\text{End}})\left[\frac{(t - t_{\text{Impact}})^2}{(t_{\text{End}} - t_{\text{Impact}})^2} - 2\frac{t - t_{\text{Impact}}}{t_{\text{End}} - t_{\text{Impact}}}\right] + N_{\text{Impact}}\right) - \frac{\sigma^2}{2}(t - t_{\text{Impact}})$ |
| Mean value curve (i.e. the parabola) | $m_{\text{parabola}}(t) = (N_{\text{Impact}} - N_{\text{End}})\left[\frac{(t - t_{\text{Impact}})^2}{(t_{\text{End}} - t_{\text{Impact}})^2} - 2\frac{t - t_{\text{Impact}}}{t_{\text{End}} - t_{\text{Impact}}}\right] + N_{\text{Impact}}$ |
| Variance | $\sigma^2_{\text{parabola}(t)} = m^2_{\text{parabola}}(t)(e^{\sigma^2(t - t_{\text{Impact}})} - 1)$ |
| Standard deviation curve | $\sigma_{\text{parabola}}(t) = m_{\text{parabola}}(t)\sqrt{e^{\sigma^2(t - t_{\text{Impact}})} - 1}$ |
| Upper standard deviation curve | $m_{\text{parabola}}(t) + \sigma_{\text{parabola}}(t) = m_{\text{parabola}}(t)\left[1 + \sqrt{e^{\sigma^2(t - t_{\text{Impact}})} - 1}\right]$ |
| Lower standard deviation curve | $m_{\text{parabola}}(t) - \sigma_{\text{parabola}}(t) = m_{\text{parabola}}(t)\left[1 - \sqrt{e^{\sigma^2(t - t_{\text{Impact}})} - 1}\right]$ |
| All the moments, i.e. $k$th moment | $\langle P^k_{\text{parabola}}(t)\rangle = \left((N_{\text{Impact}} - N_{\text{End}})\left[\frac{(t - t_{\text{Impact}})^2}{(t_{\text{End}} - t_{\text{Impact}})^2} - 2\frac{t - t_{\text{Impact}}}{t_{\text{End}} - t_{\text{Impact}}}\right] + N_{\text{Impact}}\right)^k e^{(k^2 - k)\frac{\sigma^2}{2}(t - t_{\text{Impact}})}$ |
| Mode (= abscissa of the lognormal peak) | $n_{\text{mode}} \equiv n_{\text{peak}} = \left((N_{\text{Impact}} - N_{\text{End}})\left[\frac{(t - t_{\text{Impact}})^2}{(t_{\text{End}} - t_{\text{Impact}})^2} - 2\frac{t - t_{\text{Impact}}}{t_{\text{End}} - t_{\text{Impact}}}\right] + N_{\text{Impact}}\right)e^{-\frac{3\sigma^2(t - t_{\text{Impact}})}{2}}$ |
| Value of the mode peak | $f_{N_{\text{DEC}}(t)}(n_{\text{mode}}) = \frac{e^{\sigma^2(t - t_{\text{Impact}})}}{\sqrt{2\pi}\sigma\sqrt{t - t_{\text{Impact}}}\left((N_{\text{Impact}} - N_{\text{End}})\left[\frac{(t - t_{\text{Impact}})^2}{(t_{\text{End}} - t_{\text{Impact}})^2} - 2\frac{t - t_{\text{Impact}}}{t_{\text{End}} - t_{\text{Impact}}}\right] + N_{\text{Impact}}\right)}$ |
| Median (= fifty-fifty probability value for $N_{\text{FIX}}(t)$) | $\text{median} = m = \left((N_{\text{Impact}} - N_{\text{End}})\left[\frac{(t - t_{\text{Impact}})^2}{(t_{\text{End}} - t_{\text{Impact}})^2} - 2\frac{t - t_{\text{Impact}}}{t_{\text{End}} - t_{\text{Impact}}}\right] + N_{\text{Impact}}\right)e^{-\frac{\sigma^2(t - t_{\text{Impact}})}{2}}$ |
| Skewness | $\frac{K_3}{(K_2)^{3/2}} = [e^{\sigma^2(t - t_{\text{Impact}})} + 2]\sqrt{e^{\sigma^2(t - t_{\text{Impact}})} - 1}$ |
| Kurtosis | $\frac{K_4}{(K_2)^2} = e^{4\sigma^2(t - t_{\text{Impact}})} + 2e^{3\sigma^2(t - t_{\text{Impact}})} + 3e^{2\sigma^2(t - t_{\text{Impact}})} - 6$ |

Consequently, the upper standard deviation curve is

$$m_{\text{parabola}}(t) + \sigma_{\text{parabola}}(t) = m_{\text{parabola}}(t) \times \left[1 + \sqrt{e^{\sigma^2(t - t_{\text{Impact}})} - 1}\right] \quad (60)$$

and the lower standard deviation curve is

$$m_{\text{parabola}}(t) - \sigma_{\text{parabola}}(t) = m_{\text{parabola}}(t) \times \left[1 - \sqrt{e^{\sigma^2(t - t_{\text{Impact}})} - 1}\right]. \quad (61)$$

Table 5 shows the statistical properties of our parabolic mass extinction model.

*Example: the parabola of the K–Pg mass extinction extending 10 centuries after impact*

At this point it is natural to check our parabolic mass extinction model against the corresponding exponential (i.e. GBM-based) mass extinction model.

In order to allow for the perfect match between the two relevant plots, we shall assume that the five numeric input values given in the subsection ‘Important special cases of $m_L(t)$ (2)’ for the GBM model are numerically kept just the same for the parabolic model also.

Thus, the following Fig. 4 is obtained for the parabolic K–Pg mass extinction.

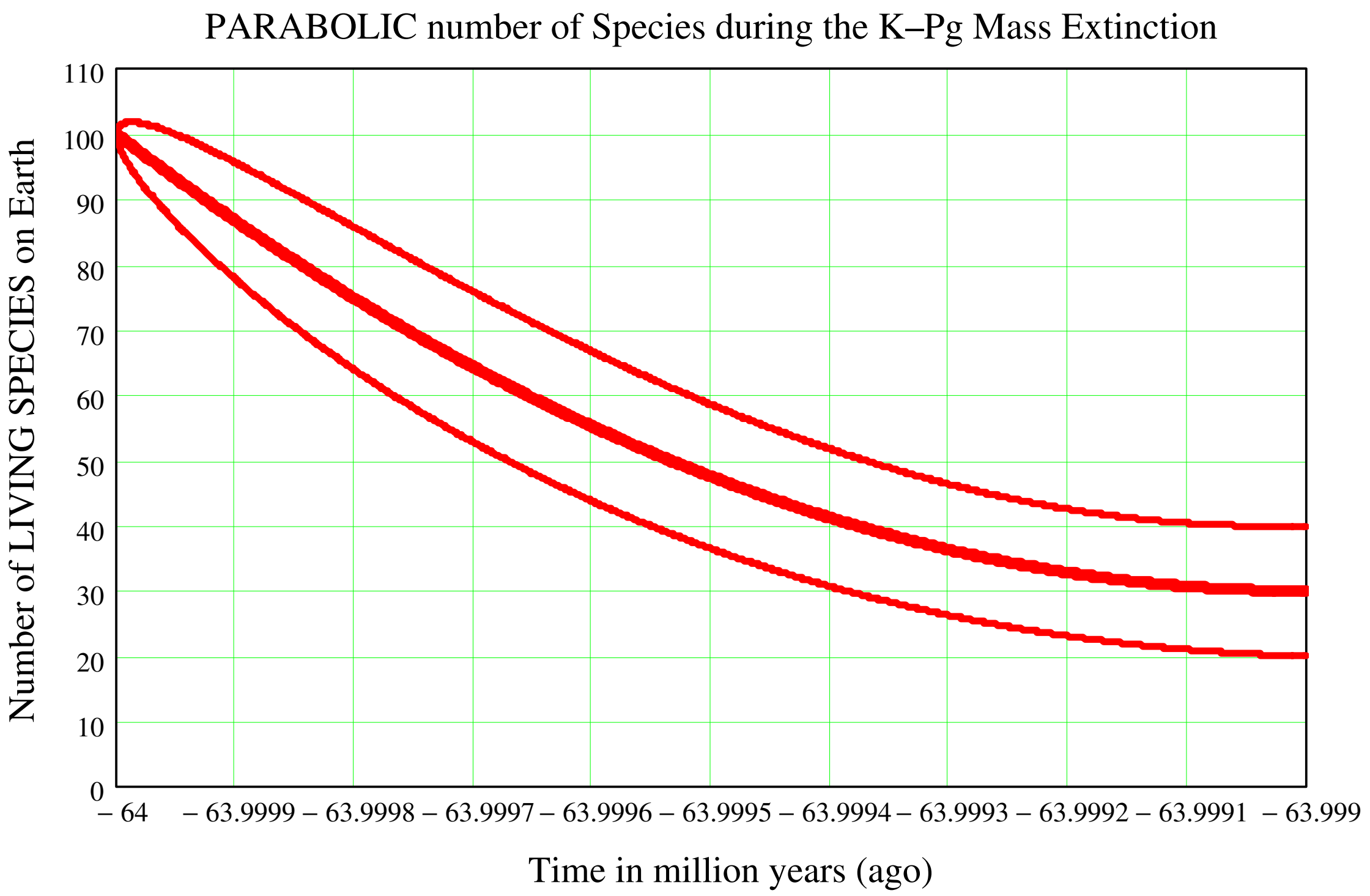

**Fig. 4.** The K–Pg mass extinction as a decreasing *parabola* in the number of living species over 1000 years after impact. The five numeric input values for this plot are just the same as those used for the construction of Fig. 3 in order to allow a perfect comparison between the two models, exponential (i.e. GBM-based) and parabolic.

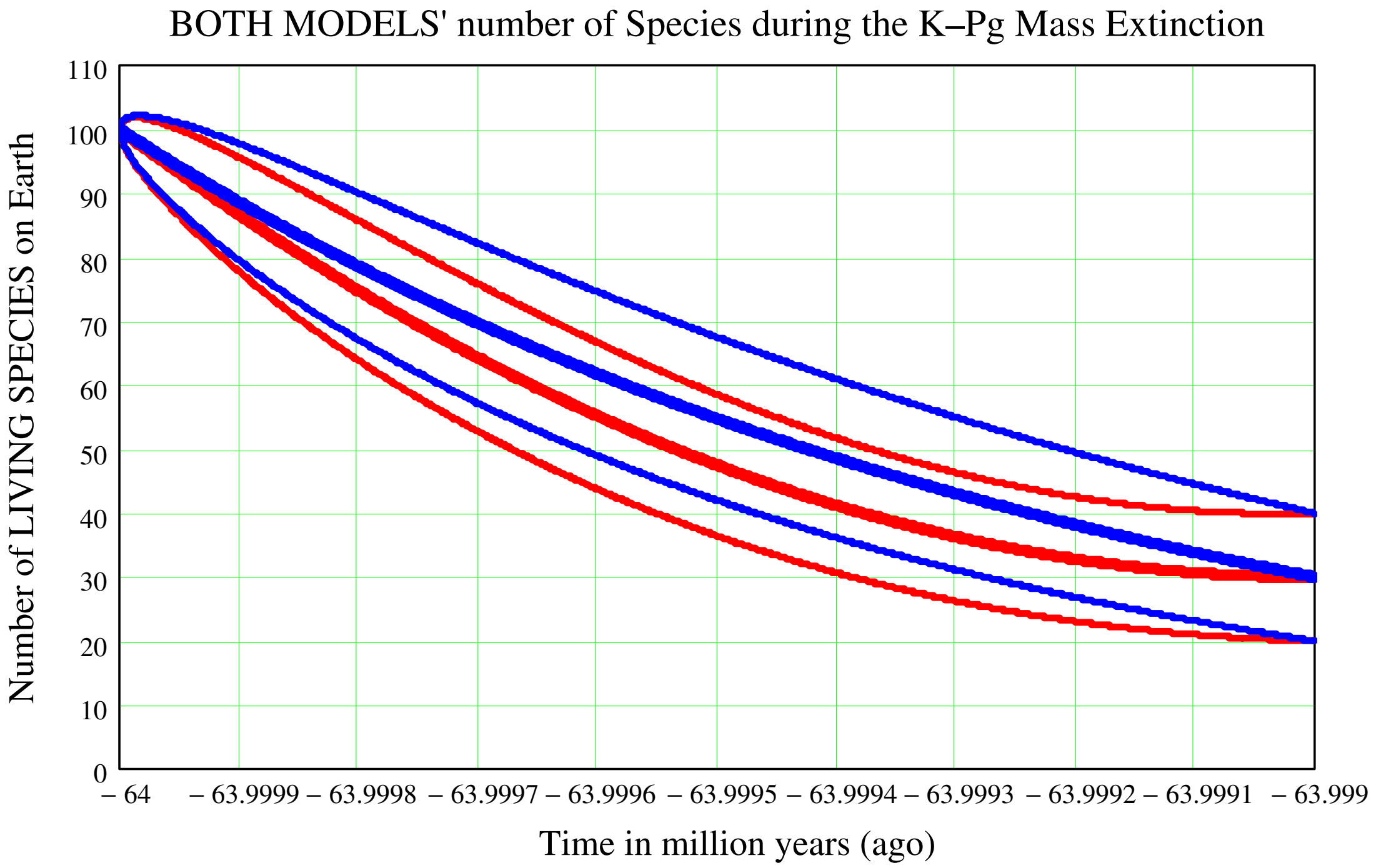

**Fig. 5.** Figs. 3 and 4 superimposed in order to allow for the perfect comparison between the two models (exponential (i.e. GBM-based) and parabolic) of the K–Pg mass extinction as a decreasing lognormal stochastic process in the number of living species over 1000 years after impact.

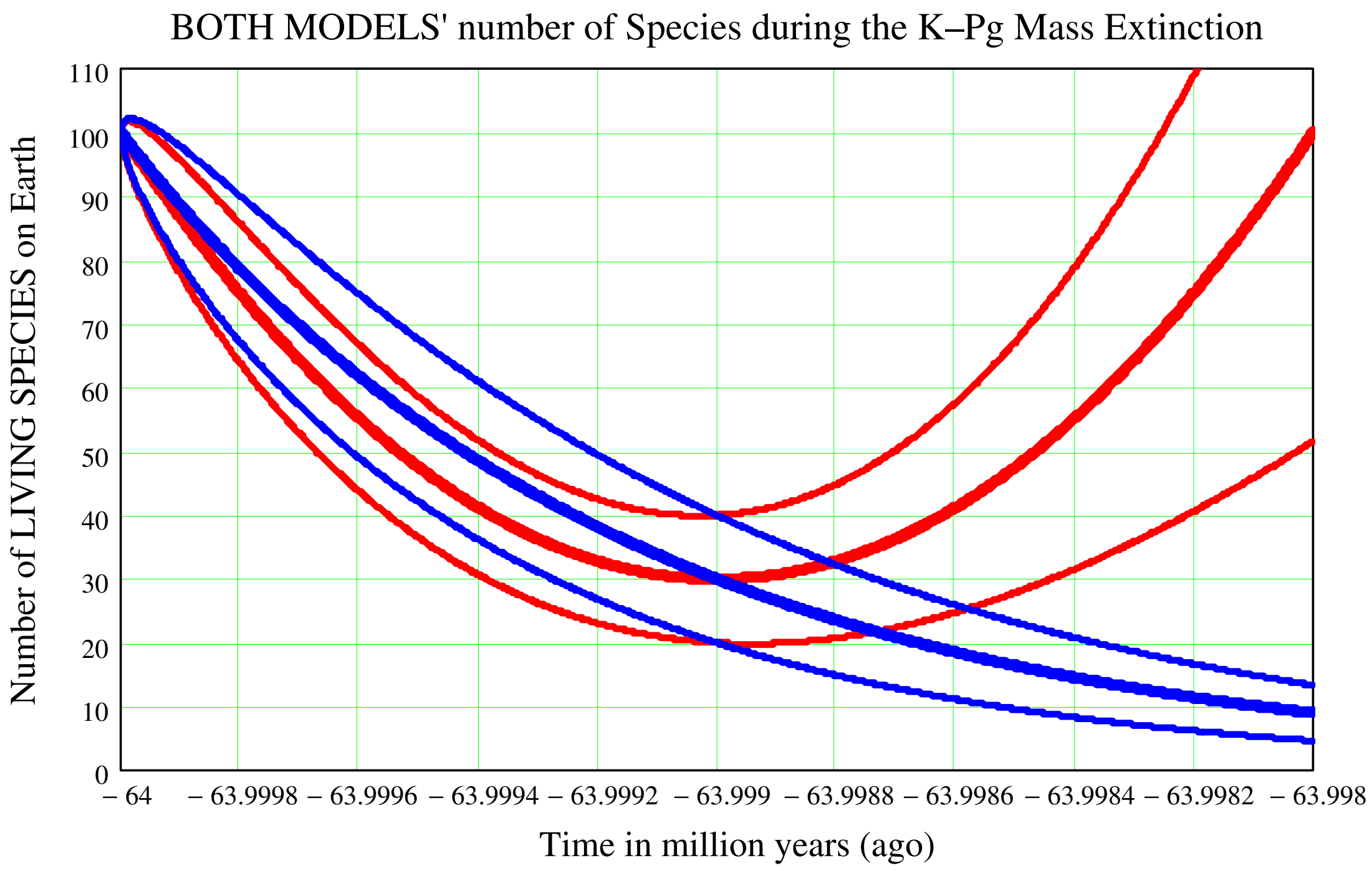

**Fig. 6.** If we *double* the horizontal axis time window of Fig. 5, then the result is the current Fig. 6. It clearly shows that the parabolic model (in red) allows for the recovery of life on Earth after the nuclear winter, while the GBM does not so. Thus, the parabolic lognormal process is a better model than the decreasing exponential (GBM) process.

Actually, we may now *superimpose* the two plots given by Figs. 3 and 4, respectively, thus obtaining Fig. 5.

## Cubic as the mean value of a lognormal stochastic process

*Finding the cubic when its maximum and minimum times are given, in addition to the five conditions to find the parabola*

Having completely solved the problem of deriving the equations of the lognormal process $L(t)$ for which the mean value is an assigned parabola, the next step is to derive the cubic (i.e. the third-degree polynomial in $t$) now assumed to be the mean value of the lognormal process $L(t)$. The relevant calculations are longer than for the parabola case, but still manageable. Unfortunately, similar calculations turn out to be too long and complicated for even higher polynomials like a quartic or a quintic: namely, analytic solutions appear to be prohibitive for polynomials higher than the cubic considered in this section, and we shall not describe here our failed attempts in this regard.

Let us start by writing down the cubic beginning with the starting time $ts$:

$$\text{Cubic}(t) = a(t-ts)^3 + b(t-ts)^2 + c(t-ts) + d. \tag{62}$$

We must determine the cubic's four coefficients ($a, b, c, d$) in terms of the following *seven inputs:*

(1) the initial time (starting time) $ts$;

(2) the initial numeric value $Ns$ of the stochastic process $L(t)$ at $ts$, namely $L_{\text{cubic}}(ts) = Ns$. To be precise, we assume that it is *certain* (i.e. with probability 1) that the process $L(t)$ will take up the value $Ns$ at the initial time $t = ts$, and so will its mean value, with a zero standard deviation there;

(3) the final time (ending time) $te$ of our lognormal $L(t)$ stochastic process;

(4) the final numeric value $Ne$ *of the mean value* of the stochastic process $L(t)$ at $te$, namely we define

$$\langle L_{\text{cubic}}(te)\rangle = Ne; \tag{63}$$

(5) in addition to the assumption (63), we also must assume that $L(t)$ will have a certain standard deviation $\delta Ne$ above and below the mean value (63) at the end-time $t = te$. These first five inputs are just the same as the five inputs described in the subsection 'Important special cases of $m_L(t)$ (1)' for the GBMs, and in the subsection 'Example: the parabola of the K–Pg mass extinction extending ten centuries after impact' for the parabola model, that in both cases we then used to describe the mass extinctions as stochastic lognormal processes.

For the cubic case we introduce two more inputs:

(6) the time of the cubic's maximum, $t_Max$; and

(7) the time of the cubic's minimum, $t_min$.

It is intuitively clear that, in order to handle the four-coefficient cubic (62), more conditions are necessary than just the previous five conditions, necessary to handle both the two-parameter GBM (9) and the three-coefficient parabola (50). However, it was not initially obvious to this author *how many* more conditions would have been necessary and especially-*which ones.* The answers to these two questions came out only by doing the actual calculations, as we now describe for the particular case when the two conditions (6) and (7) reveal themselves sufficient to determine the cubic (62) completely. This particular way of determining the cubic is important in the study of Darwinian Evolution as described by the contemporary Russian scientist Andrey Korotayev and his colleague Alexander V. Markov, which we shall study in the next section.

Going over to the actual calculations, we notice that, because of the two initial conditions (1) and (2), the cubic (62) yields, respectively

$$\begin{cases} \text{Cubic}(ts) = d, \\ \text{Cubic}(ts) = Ns. \end{cases} \tag{64}$$

These, inserted into the cubic (62), change to

$$\text{Cubic}(t) = a(t - ts)^3 + b(t - ts)^2 + c(t - ts) + Ns. \tag{65}$$

We then invoke the two final conditions (3) and (4) that translate into the single equation

$$\text{Cubic }(te) = Ne. \tag{66}$$

In other words, (66) changes the cubic (62) to

$$Ne - Ns = a(te - ts)^3 + b(te - ts)^2 + c(te - ts). \tag{67}$$

The only three unknowns in (67) are the three still unknown cubic coefficients $(a, b, c)$. But actually (67) is a relationship among these three coefficients $(a, b, c)$. Thus, in reality, we only need *two* more conditions yielding, for instance, both $b$ and $c$ as functions of $a$, respectively, and we would then insert them both into (67) that would then become an equation with the only unknown $a$. Solving that equation for $a$ would then solve the problem completely, yielding then both $b$ and $c$ as functions of all known quantities. So, let us now look for these *two* still missing conditions on $(a, b, c)$. To this end, the key idea is that every cubic has both a maximum and a minimum. To find them, the cubic's (62) first derivative with respect to $t$ must be set equal to zero:

$$\frac{\text{dCubic}(t)}{\text{d}t} = 3a(t - ts)^2 + 2b(t - ts) + c = 0. \tag{68}$$

Solving this quadratic for $t$ yields the two roots:

$$\begin{cases} t_1 = ts + \dfrac{-b - \sqrt{b^2 - 3ac}}{3a} = ts + X_1, \\ t_2 = ts + \dfrac{-b + \sqrt{b^2 - 3ac}}{3a} = ts + X_2, \end{cases} \tag{69}$$

having set

$$\begin{cases} X_1 = \dfrac{-b - \sqrt{b^2 - 3ac}}{3a}, \\ X_2 = \dfrac{-b + \sqrt{b^2 - 3ac}}{3a}. \end{cases} \tag{70}$$

Note that the two equations (69) yield the abscissas (i.e. the instants) of the two stationary points of the quadratic (68), but we do not know which ones, i.e. we do not know which one is the maximum and which one is the minimum. If we suppose that the abscissas (i.e. the instants) of the maximum and the minimum of the cubic (62) are assigned, i.e. they are known, then $X_1$ *and* $X_2$ *are also known*, since they are the same as the maximum and the minimum except for the additional time $ts$, the starting time of the cubic (62). By doing so, we have indeed taken the two conditions (6) and (7) into account.

Adding the equations in (70) and then solving for $b$ yields

$$b = -\frac{3a(X_1 + X_2)}{2}, \tag{71}$$

that is the expression of $b$ as a function of $a$ that we were seeking. Similarly, multiplying the equations in (70) and then solving for $c$ yields the required expression of $c$ as a function of $a$:

$$c = 3aX_1X_2. \tag{72}$$

So, by substituting the two equations (71) and (72) into (67), we get an equation in the only unknown $a$ that is

$$Ne - Ns = (te - ts) \times \left[a(te - ts)^2 - \frac{3a(X_1 + X_2)}{2}(te - ts) + 3aX_1X_2\right]. \tag{73}$$

Solving (73) for $a$ yields

$$a = \frac{2(Ne - Ns)}{(te - ts)[2(te - ts)^2 - 3(X_1 + X_2)(te - ts) + 6X_1X_2]}. \tag{74}$$

Next we find $b$ by substituting (74) into (71)

$$b = \frac{-3(X_1 + X_2)(Ne - Ns)}{(te - ts)[2(te - ts)^2 - 3(X_1 + X_2)(te - ts) + 6X_1X_2]}, \tag{75}$$

and we also find $c$ by substituting (74) into (72)

$$c = \frac{6X_1X_2(Ne - Ns)}{(te - ts)[2(te - ts)^2 - 3(X_1 + X_2)(te - ts) + 6X_1X_2]}. \tag{76}$$

Thus, the cubic (65) is now obtained by substituting (74)–(76) into (65), with the result

$$\text{Cubic}(t) = (Ne - Ns) \times \frac{(t - ts)[2(t - ts)^2 - 3(X_1 + X_2)(t - ts) + 6X_1X_2]}{(te - ts)[2(te - ts)^2 - 3(X_1 + X_2)(te - ts) + 6X_1X_2]} + Ns. \tag{77}$$

A glance to (77) immediately reveals that both the boundary conditions given by the lower equations (64) and (66), respectively, and indeed fulfilled. But the important point is to notice that the cubic (77) is *symmetric* in $X_1$ and $X_2$, namely that the cubic (77) does not change at all if $X_1$ and $X_2$ are interchanged. Again, this is another way to say that 'we do not know which one, out of $X_1$ and $X_2$, corresponds to the abscissa of the maximum and the abscissa of the minimum'. The answer to this apparent 'surprise' is that it all depends on the factor $(Ne–Ns)$ in front of the fraction in (77):

(1) if $Ne > Ns$ then the cubic's coefficient of $t^3$ in (77) is *positive*. Then, the cubic 'starts' at $-\infty$, grows up to its maximum, then goes down to its minimum, and finally 'climbs up again on the right'. In other words, the maximum is reached before the minimum. And this will be the case of the Markov–Korotayev's cubic of evolution that we shall study in the next section.

(2) if $Ne < Ns$, it is the other way round. That is, the cubic 'starts' at $+\infty$, gets down to its minimum first, then it climbs up to its maximum, and finally gets down to $-\infty$ on the right. In other words, its minimum comes before its maximum.

But there is still a better form of (77) that we wish to point out. This comes from the replacement of $X_1$ and $X_2$ in (77) by virtue of the explicit abscissa of the maximum, $t$_Max, and of the minimum, $t$_min, related to $X_1$ and $X_2$ via (69), that is (assuming for simplicity that $Ne > Ns$, as in the Markov–Korotayev case):

$$\begin{cases} t_{\mathrm{Max}} = ts + \dfrac{-b - \sqrt{b^2 - 3ac}}{3a} = ts + X_1, \\ t_{\mathrm{min}} = ts + \dfrac{-b + \sqrt{b^2 - 3ac}}{3a} = ts + X_2, \end{cases} \tag{78}$$

from which one gets

$$\begin{cases} X_1 = t_{\mathrm{Max}} - ts, \\ X_2 = t_{\mathrm{min}} - ts. \end{cases} \tag{79}$$

Thus, inserting (79) into (77), we reach our final version of the cubic mean value of the $L(t)$ lognormal stochastic process

$$\mathrm{Cubic}(t) = (Ne - Ns)\frac{(t-ts)\,[2(t-ts)^2 - 3(t_{\mathrm{Max}} + t_{\mathrm{min}} - 2ts)(t-ts) + 6(t_{\mathrm{Max}} - ts)(t_{\mathrm{min}} - ts)]}{(te-ts)\,[2(te-ts)^2 - 3(t_{\mathrm{Max}} + t_{\mathrm{min}} - 2ts)(te-ts) + 6(t_{\mathrm{Max}} - ts)(t_{\mathrm{min}} - ts)]} + Ns. \tag{80}$$

Our reader might have noticed that the condition (5) was *not* used to derive the cubic (80). This is because the condition (5) does not affect the cubic (80): it only affects the standard deviation $\sigma_{\mathrm{Cubic}}(t)$ and the two corresponding upper and lower standard deviation curves above and below the mean value cubic (80). This fact is evident from equation (3) clearly showing that the time function $M_L(t)$ and the positive parameter $\sigma_L$ have nothing to do with each other, i.e. they are independent of each other, just as the mean value and the variance of the Gaussian (normal) distribution are totally independent of each other. Thus, going back to equation (39), we conclude that it does not hold good for GBMs only, but rather it applies to all lognormal stochastic processes, whatever their mean value $m_L(t)$ might possibly be. In conclusion, the positive parameter σ is determined by (39) just rewritten in this section's notation:

$$\sigma = \sqrt{\frac{\ln\left[1 + \left(\dfrac{\delta Ne}{Ne}\right)^2\right]}{te - ts}}. \tag{81}$$

We are now ready to write down the two equations of the upper and lower standard deviation curves. They are actually the same as the two equations at the seventh and eighth lines in Table 1, which we re-write here in the current 'cubic' notation:

$$\begin{cases} \text{upper_standard_deviation_curve}(t) = m_{\mathrm{Cubic}(t)} + \sigma_{\mathrm{Cubic}(t)} \\ \quad = \mathrm{Cubic}(t)\left[1 + \sqrt{e^{\sigma^2(t-ts)} - 1}\right], \\ \text{lower_standard_deviation_curve}(t) = m_{\mathrm{Cubic}(t)} - \sigma_{\mathrm{Cubic}(t)} \\ \quad = \mathrm{Cubic}(t)\left[1 - \sqrt{e^{\sigma^2(t-ts)} - 1}\right]. \end{cases} \tag{82}$$

## Markov–Korotayev biodiversity regarded as a lognormal stochastic process having a cubic mean value

### *Markov–Korotayev's work on evolution*

Let us now refer to the interesting Wikipedia site http://en.wikipedia.org/wiki/Andrey_Korotayev, whom we quote verbatim. According to this, in 2007–2008 the Russian scientist Andrey Korotayev, in collaboration with Alexander V. Markov showed that a 'hyperbolic' mathematical model can be developed to describe the macrotrends of biological evolution. These authors demonstrated that changes in biodiversity through the Phanerozoic correlate much better with the hyperbolic model (widely used in demography and macrosociology) than with the exponential and logistic models (traditionally used in population biology and extensively applied to fossil biodiversity as well). The latter models imply that changes in diversity are guided by a first-order positive feedback (more ancestors, more descendants) and/or a negative feedback arising from resource limitation. Hyperbolic model implies a second-order positive feedback. The hyperbolic pattern of the world population growth has been demonstrated by Korotayev to arise from a second-order positive feedback between the population size and the rate of technological growth. According to Korotayev and Markov, the hyperbolic character of biodiversity growth can be similarly accounted for by a feedback between the diversity and community structure complexity. They suggest that the similarity between the curves of biodiversity and human population probably comes from the fact that both are derived from the interference of the hyperbolic trend with cyclical and stochastic dynamics (Markov & Korotayev 2007, 2008).

This author was astounded by Fig. 7 (taken from Wikipedia) showing the increase, *but not monotonic increase*, of the number of genera (in thousands) during the 542 million years making up for the Phanerozoic. Thus, this author came to wonder whether the red curve in Fig. 7 could be regarded as the cubic mean value curve of a lognormal stochastic process, just as the exponential mean value curve is typical of GBMs.

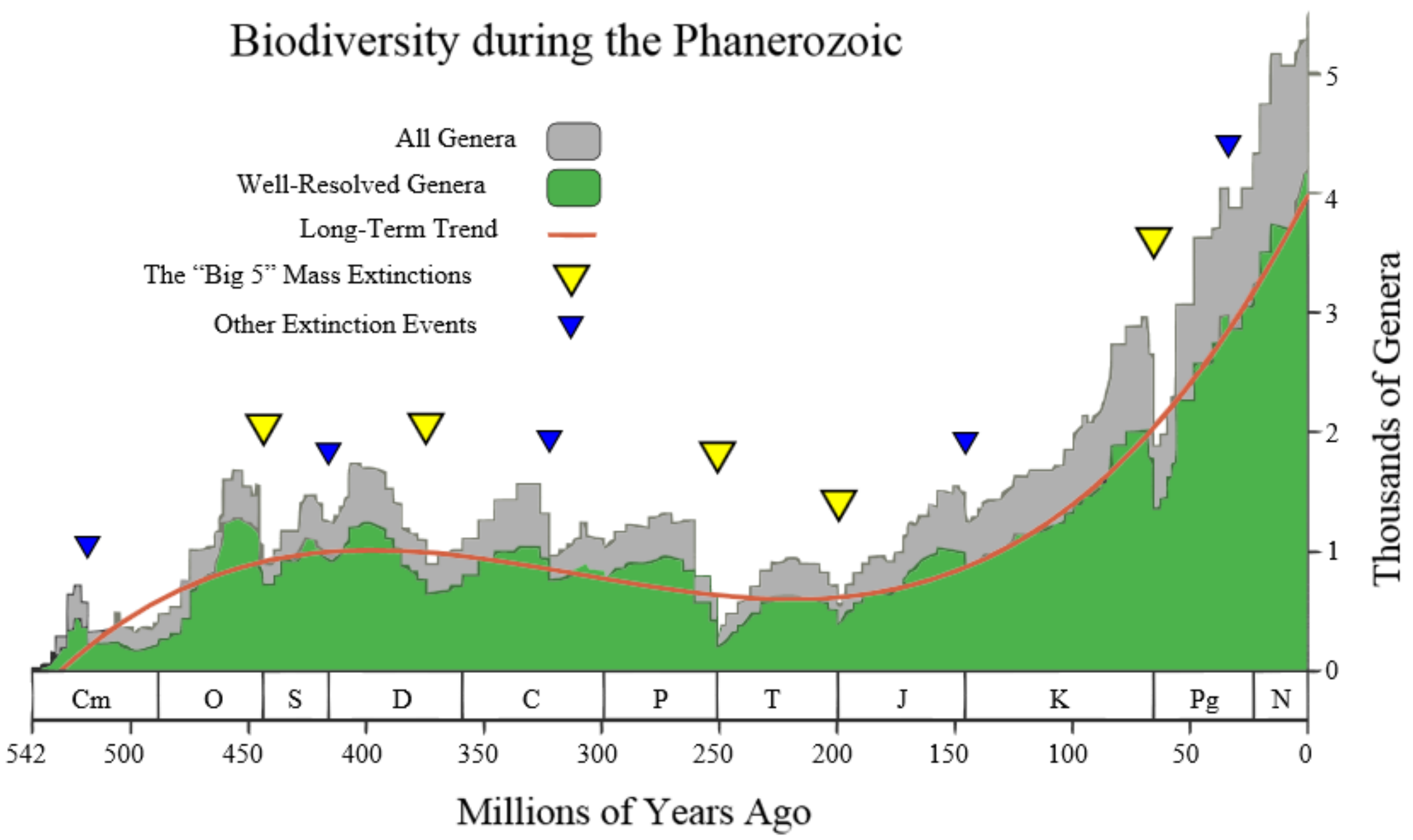

**Fig. 7.** During the Phanerozoic the biodiversity shows a steady but not monotonic increase from near zero to several thousands of genera.

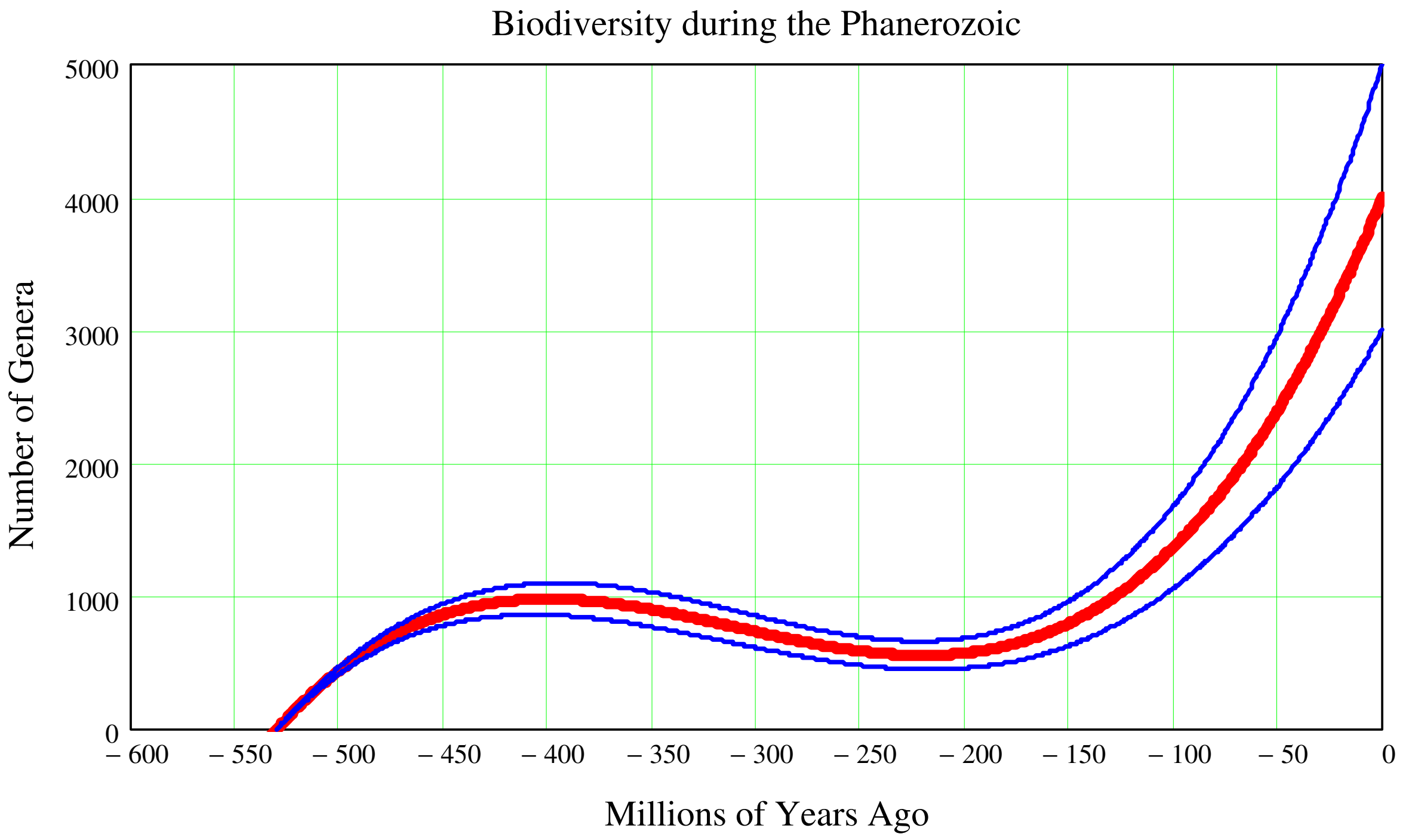

**Fig. 8.** Our cubic mean value curve (thick red solid curve) plus and minus the two standard deviation curves (thin solid blue curves) give more mathematical information than just the previous Fig. 7. In fact, we now have the two standard deviation curves of the lognormal stochastic process $L(t)$ that are completely missing in the Markov–Korotayev theory and in their plot shown in Fig. 7. We thus claim that our cubic mathematical theory of the lognormal stochastic process $L(t)$ is a 'more profound mathematization' than the Markov–Korotayev theory of evolution since it is stochastic, rather than just deterministic. This completes our 'stochastic extension' of the Markov–Korotayev evolution model.

This author's answer to the above question is 'yes': we may indeed use our cubic (80) to represent the red line in Fig. 5, thus reconciling the Markov–Korotayev theory with our theory requiring that the profile curve of evolution must be the cubic mean value curve of a certain lognormal stochastic process (and certainly *not* a GBM in this case). Let us thus consider the following numerical inputs to the cubic (80) that we derive 'by a glance to Fig. 7' (the precision of these numerical inputs is really unimportant at this early stage of 'matching' the two theories, ours and the Markov–Korotayev's, since we are just looking for the 'proof of concept', and better numeric approximations might follow in the future):

$$\begin{cases} ts = -530, \\ Ns = 1, \\ te = 0, \\ Ne = 4000. \end{cases} \tag{83}$$

In other words, the first two equations in (83) mean that the first of the genera appeared on Earth about 530 million years ago, i.e. before that time the number of genera on Earth was zero. Also, the last two equations in (83) mean that at the present time $t=0$, the number of genera on Earth is 4000 *on average*. Now, 'on average' means that, nowadays, a standard deviation of about 1000 (plus or minus) affects the average value of 4000. This is shown in Fig. 7 by the grey stochastic process called 'all genera'. And this is re-phrased mathematically by invoking the condition (5) of subsection 'Finding the cubic when its maximum and minimum times are given, in addition to the five conditions to find the parabola', and assigning the fifth numeric input

$$\delta Ne = 1000. \tag{84}$$

Then, as a consequence of the four numeric boundary inputs (83) plus the standard deviation on the current value of genera (84), equation (81) yields the numeric value of the positive parameter σ,

$$\sigma = \sqrt{\frac{\ln\left[1 + \left(\frac{\delta Ne}{Ne}\right)^2\right]}{te - ts}} = 0.011. \tag{85}$$

Having thus assigned numeric values to the first five conditions of the subsection 'Finding the cubic when its maximum and minimum times are given, in addition to the five conditions to find the parabola', only conditions (6) and (7) remain to be assigned. These are the two abscissae of the maximum and minimum, respectively, which at a glance at Fig. 7 makes us establish as (of course in millions of years ago)

$$\begin{cases} t_{Max} = -400, \\ t_{min} = -220. \end{cases} \tag{86}$$

Inserting these seven numeric inputs into the cubic (80) and into both the equations (82) of the upper and lower standard deviation curves, the final plot shown in Fig. 8 is produced.

## Conclusions

Let us finally reach the conclusions of this paper:

(1) In section 'A summary of the 'Evo-SETI' model of evolution and SETI' we showed how to 'construct' a lognormal stochastic process $L(t)$ whose mean is an assigned and 'arbitrary' function of the time $m_L(t)$. This is of paramount importance for all future applications.

(2) In the practice, this 'arbitrary' mean time $m_L(t)$ may be either an exponential $N_0 e^{\mu t}$, and then the corresponding lognormal process $L(t)$ is the well-known GBM, or it may be a polynomial function of the time, $\sum_{k=0}^{\text{polynomial_degree}} c_k t^k$, and then we have shown how to compute all the statistical properties of the corresponding lognormal process $L(t)$.

(3) In particular, we have given an important application of this duality (either exponential or polynomial assumed as mean value) in the case of the mass extinctions that plagued the development of life on Earth several times over the last 3.5 billion years. Our result is that the parabolic model is preferable to the GBM model for mass extinctions, inasmuch as the possibility of the recovery of life (as indeed it always happened on Earth) is in the parabolic model, but *not* in the GBM one.

(4) Finally, we compared our last 'stochastic cubic' result with the Markov–Korotayev model of evolution of life on Earth based on a cubic-shaped mean value function for $L(t)$. We conclude that their model and ours agree quite well, but ours is 'mathematically more profound' since it also provides both upper and lower standard deviation curves that are not present in the Markov–Korotayev model since it is deterministic, rather than stochastic, like ours.

In conclusion, we have uncovered an important generalization of GBMs into a lognormal stochastic process $L(t)$ having an *arbitrary* mean, rather than just an exponential one.

These results should pave the way for a future understanding of the evolution of life on *exoplanets* on the basis of the way the evolution of life unfolded on Earth over the last 3.5 billion years.

That will be the goal of our research papers.

## Supplementary material

Supplementary materials of this paper is available at http://dx.doi.org/10.1017/S147355041400010X